\documentclass[preprintnumbers,amsmath,amssymb,floatfix,11
pt,prd,superscriptaddress,nofootinbib]{revtex4}
\usepackage{graphicx}
\usepackage{epsfig}
\usepackage{bm}
\usepackage{amsfonts}

\input epsf

\begin{document}

\title{\Large Fractional Action Cosmology: Some Dark Energy Models in Emergent, Logamediate and Intermediate Scenarios of the Universe}

\author{Ujjal Debnath}
\email{ujjaldebnath@yahoo.com , ujjal@iucaa.ernet.in}
\affiliation{Department of Mathematics, Bengal Engineering and
Science University, Shibpur, Howrah-711 103, India.}

\author{Surajit Chattopadhyay}
\email{surajit_2008@yahoo.co.in,
surajit.chattopadhyay@pcmt-india.net} \affiliation{Department of
Computer Application (Mathematics Section), Pailan College of
Management and Technology, Bengal Pailan Park, Kolkata-700 104,
India.}

\author{Mubasher Jamil}
\email{mjamil@camp.nust.edu.pk , jamil.camp@gmail.com}
\affiliation{Center for Advanced Mathematics and Physics (CAMP),
National University of Sciences and Technology (NUST), H-12,
Islamabad, Pakistan.}

\begin{abstract}
\vspace*{1.5cm} \centerline{\bf Abstract} \vspace*{1cm} In the
framework of Fractional Action Cosmology, we have reconstructed
the scalar potentials and scalar fields, namely, quintessence,
phantom, tachyon, k-essence, DBI-essence, Hessence, dilaton field
and Yang-Mills field. To get more physical picture of the
variation of the scalar field and potential with time, we express
scale factor in emergent, logamediate and intermediate scenarios,
under which the Universe expands differently.
\end{abstract}


\maketitle

\newpage

\section{Introduction}

Fractional action cosmology (FAC) is based on the principles and
formalism of the fractional calculus applied to cosmology. The
fractional derivative and fractional integrals are the main tools
in fractional calculus, where the order of differentiation or
integration is not an integer. The fractional calculus is
immensely useful in various branches of mathematics, physics and
engineering \cite{rami}. In doing FAC, one can proceed in two
different ways \cite{mark}: the first one is quite easy as one has
to replace the partial derivatives in the Einstein field equations
with the corresponding fractional derivatives; the second
technique involves deriving the field equations and geodesic
equations from a more fundamental way, namely starting with the
principle of least action and replacing the usual integral with a
fractional integral. This later technique is more useful in giving
extra features of the FAC \cite{Rami}: Rami introduced the FAC by
introducing the fractional time integral,
\begin{equation}
S=-\frac{m}{2\Gamma(\xi)}\int\dot{x}^\mu\dot{x}^\nu
g_{\mu\nu}(x)(t-\tau)^{\xi-1}d\tau.\tag{Ia}
\end{equation}
Here $\Gamma(\xi)=\int_0^\infty t^{\xi-1}e^{-t}dt$ is the Gamma
function, $0<\xi\leq1$, $0<\tau <t$, $m=$ constant and $\dot
x^\mu=\frac{dx^\mu}{d\tau}$. The variation yields an extra term in
the field equations which he termed as `variable gravitational
constant $G$'. Moreover, when the weight function in the
fractional time integral is replaced with a sinusoidal function,
then the solution of the corresponding field equations yield a
variable cosmological constant and an oscillatory scale factor
\cite{rami1}
\begin{equation}
S=\frac{m}{2}\int_0^\tau \dot{x}^\mu\dot{x}^\nu
g_{\mu\nu}(x)e^{-\chi \sin(\beta t)}dt,\tag{Ib}
\end{equation}
where $\chi=0$ reduces to the standard action. In \cite{jamil},
the authors extended the previous study by working out with a
general weight function:
\begin{equation} S=\frac{m}{2}\int\limits_0^\tau
g_{\mu\nu}(x)\dot{x}^{\mu}\dot{x}^{\nu}\mu(\chi,t)dt,\tag{Ic}
\end{equation}
 Several examples were studied and cosmological
parameters were calculated in there. An interesting feature of FAC
is that it yields an expanding Universe whose scale factor goes like
power law form or exponential form depending on the choice of the
weight function. Hence cosmic acceleration can be modeled in FAC.

Reconstruction of potentials has been done by several authors in
various cases. Capozziello et al \cite{Capozziello} considered
scalar-tensor theories and reconstruct their potential and
coupling by demanding a background $\Lambda$CDM cosmology. In the
framework of phantom quintessence cosmology, \cite{Capozziello1}
used the Noether Symmetry Approach to obtain general exact
solutions for the cosmological equations.In this paper, we are
going to reconstruct the potentials and scalar fields, namely,
quintessence, phantom, tachyonic, k-essence, DBI-essence,
Hessence, dilaton field and Yang-Mills field. Such reconstructions
have been studied previously in other gravitational setups
\cite{other}. To get more physical insight into the model, we
express scale factor in three useful forms \cite{obs} namely
emergent, logamediate and intermediate scenarios, under which the
Universe expands differently. Such expansion scenarios are
consistent with the observations with some restrictions on their
parameters \cite{obs}.

\section{Fractional Action Cosmological Model}

For a FRW spacetime, the line element is
\begin{equation}
ds^{2} = - dt^{2} + a^{2}(t) \left[\frac{ dr^{2}}{1-k r^{2}} +
r^{2}(d\theta^{2}+\sin^{2}\theta d\phi^{2})\right],
\end{equation}
where $a(t)$ is the scale factor and $k ~(= 0, \pm 1)$ is the
curvature scalar. We consider the Universe contains normal matter
and dark energy. From Eq. (Ia), the Einstein equations for the
space-time given by equation (1) are \cite{Rami}
\begin{eqnarray}
H^{2}+\frac{2(\xi-1)}{T_1}H+\frac{k}{a^{2}}&=&\frac{8\pi
G}{3}\rho,\\
\dot{H}-\frac{(\xi-1)}{T_1}H-\frac{k}{a^{2}}&=& -4\pi G (\rho+p),
\end{eqnarray}
where $T_1=t-\tau$, $\rho=(\rho_{m}+\rho_{\phi})$ and
$p=(p_{m}+p_{\phi})$. Here $\rho_{m}$ and $p_{m}$ are the energy
density and pressure of the normal matter connected by the equation
of state
\begin{equation}
p_{m}=w_{m}\rho_{m}~,~~-1\le w_{m}\le 1
\end{equation}
and $\rho_{\phi}$ and $p_{\phi}$ are the energy density and
pressure due to the dark energy.

Now consider there is an interaction between normal matter and
dark energy. Dark energy interacting with dark matter is a
promising model to alleviate the cosmic coincidence problem. In
Ref. \cite{wang1}, the authors studied the signature of such
interaction on large scale cosmic microwave background (CMB)
temperature anisotropies. Based on the detail analysis in
perturbation equations of dark energy and dark matter when they
are in interaction, they found that the large scale CMB,
especially the late Integrated Sachs Wolfe effect, is a useful
tool to measure the coupling between dark sectors. It was deduced
that in the 1$\sigma$ range, the constrained coupling between dark
sectors can solve the coincidence problem. In Ref. \cite{wang2}, a
general formalism to study the growth of dark matter perturbations
when dark energy perturbations and interactions between dark
sectors were presented. They showed that the dynamical stability
on the growth of structure depends on the form of coupling between
dark sectors. Moreover due to the influence of the interaction,
the growth index can differ from the value without interaction by
an amount up to the observational sensibility, which provides an
opportunity to probe the interaction between dark sectors through
future observations on the growth of structure.

Due to this interaction, the normal matter and dark energy are not
separately conserved. The energy conservation equations for normal
matter and dark energy are
\begin{equation}
\dot{\rho}_{m} + 3 H( p_{m} + \rho_{m} ) = -3\delta H\rho_{m},
\end{equation}
and
\begin{equation}
\dot{\rho}_{\phi} + 3 H( p_{\phi} + \rho_{\phi} ) = 3\delta
H\rho_{m},
\end{equation}
where $H=\dot{a}/a$ is the Hubble parameter.

From equation (5) we have the expression for energy density of
matter as
\begin{equation}
\rho_{m} = \rho_{0} a^{-3 ( 1+ w_{m}+\delta)},
\end{equation}
where $ \rho_{0} $ is the integration constant.

\section{Emergent, Logamediate and Intermediate Scenarios}

$\bullet${\bf Emergent Scenario}: For emergent Universe, the scale
factor can be chosen as \cite{debnath}
\begin{equation}
a(T_1) = a_{0}\left(\lambda+e^{\mu T_1}\right)^{n}
\end{equation}
where $a_{0},~\mu,~\lambda$ and $n$ are positive constants. (1)
$a_0
> 0$ for the scale factor $a$ to be positive; (2) $\lambda > 0$, to
avoid any singularity at finite time (big-rip); (3) $a > 0$ or $n >
0$ for expanding model of the Universe; (4) $a < 0$ and $n < 0$
implies big bang singularity at $t = -\infty$.

 So the
Hubble parameter and its derivatives are given by
\begin{equation}
H=\frac{n\mu e^{\mu T_1}}{\left(\lambda+e^{\mu T_1}\right)}~,~
\dot{H}=\frac{n\lambda\mu^{2}e^{\mu T_1}}{\left(\lambda+e^{\mu
T_1}\right)^{2}}~,~\ddot{H}=\frac{n\lambda\mu^{3}e^{\mu
T_1}(\lambda-e^{\mu T_1})}{\left(\lambda+e^{\mu T_1}\right)^{3}}
\end{equation}
Here $H$ and $\dot{H}$ are both positive, but $\ddot{H}$ changes
sign at $T_1=\frac{1}{\mu}~\text{log}\lambda$. Thus $H,~\dot{H}$ and
$\ddot{H}$ all tend to zero as $t\rightarrow -\infty$. On the other
hand as $t\rightarrow \infty$ the solution gives
asymptotically a de Sitter Universe.\\

$\bullet${\bf Logamediate Scenario}: Consider a particular form of
Logamediate Scenario, where the form of the scale factor $a(t)$ is
defined as \cite{obs}
\begin{equation}
a(T_1)=e^{A(\ln T_1)^{\alpha}},
\end{equation}
where $A \alpha>0$ and $\alpha>1$. When $\alpha=1$, this model
reduces to power-law form. The logamediate form is motivated by
considering a class of possible cosmological solutions with
indefinite expansion which result from imposing weak general
conditions on the cosmological model. Barrow has found in their
model, the observational ranges of the parameters are as follows:
$1.5\times 10^{-92}\le A \le 2.1\times 10^{-2}$ and $2\le \alpha
\le 50$. The Hubble parameter $H=\frac{\dot{a}}{a}$ and its
derivative become,
\begin{equation}
H=\frac{A\alpha}{T_1}(\ln
T_1)^{\alpha-1}~~,~~\dot{H}=\frac{A\alpha}{T_1^{2}}(\ln
T_1)^{\alpha-2}(\alpha-1-\ln T_1)
\end{equation}

$\bullet${\bf Intermediate Scenario}: Consider a particular form of
Intermediate Scenario, where the scale factor $a(t)$ of the
Friedmann universe is described as \cite{obs},
\begin{equation}
a(t)=e^{B T_1^\beta},
\end{equation}
 where $B\beta>0$, $B>0$ and $0<\beta<1$. Here the expansion of Universe is faster than Power-Law
form, where the scale factor is given as, $a(T_1) = T_1^n$, where
$n>1$ is a constant. Also, the expansion of the Universe is slower
for Standard de-Sitter Scenario where $\beta = 1$.
 The Hubble parameter $H=\frac{\dot{a}}{a}$ and its derivative become,
\begin{equation}
H=B\beta T_1^{\beta-1}~,~~\dot{H}=B\beta(\beta-1) T_1^{\beta-2}
\end{equation}

\section{Various Candidates of Dark Energy Models}

\subsection{Quintessence or Phantom field}

Quintessence is described by an ordinary time dependent and
homogeneous scalar field $\phi$ which is minimally coupled to
gravity, but with a particular potential $V(\phi)$ that leads to the
accelerating Universe. The action for quintessence is given by
\cite{Copeland}
\begin{equation}
S=\int {\rm d}^4x\sqrt{-g}\left[ -\frac{1}{2}g^{ij}\partial_i
\phi\partial_j \phi-V(\phi) \right].\nonumber
\end{equation}
The energy momentum tensor of the field is:
\begin{equation}\label{2}
T_{ij}=-\frac{2}{\sqrt{-g}}\frac{\delta S}{\delta
g^{ij}},\nonumber
\end{equation}
which gives
\begin{equation}\label{3}
T_{ij}=\partial_i \phi\partial_j \phi-g_{ij}
\left[\frac{1}{2}g^{kl}\partial_k \phi\partial_l \phi
+V(\phi)\right].\nonumber
\end{equation}
 The energy density and pressure of the quintessence scalar field
$\phi$ are as follows
\begin{equation}
\rho_\phi=-T^0_0=\frac{1}{2}\dot \phi^2+V(\phi),\label{ro
q}\nonumber
\end{equation}
\begin{equation}
p_\phi=T_i^i=\frac{1}{2}\dot \phi^2-V(\phi).\label{p q}\nonumber
\end{equation}
The EoS parameter for the quintessence scalar field is given by
\begin{equation}
\omega_\phi=\frac{p_\phi}{\rho_\phi}=\frac{\dot
\phi^2-2V(\phi)}{\dot \phi^2+2V(\phi)}.\label{w q}\nonumber
\end{equation}
For $\omega_\phi<-1/3$, we find that the Universe accelerates when
$\dot \phi^2<V(\phi).$

The energy density and the pressure of the quintessence (phantom
field) can be represented by the minimally coupled spatially
homogeneous and time dependent scalar field $\phi$ having positive
(negative) kinetic energy term given by
\begin{equation}
\rho_{\phi}=\frac{\epsilon}{2}~\dot{\phi}^{2}+V(\phi)
\end{equation}
and
\begin{equation}
p_{\phi}=\frac{\epsilon}{2}~\dot{\phi}^{2}-V(\phi)
\end{equation}
where $V(\phi)$ is the relevant potential for the scalar field
$\phi$, $\epsilon=+1$ represents quintessence while $\epsilon=-1$
refers to phantom field.

Scalar field models of phantom energy indicate that it can behave
as a long range repulsive force \cite{Luca}. Moreover the phantom
energy has few characteristics different from normal matter, for
instance, the energy density $\rho(t)$ of the phantom field
increases with the expansion of the Universe; it can be used as a
source to form and stabilize traversable wormholes
\cite{ellis,picon,rahm,kuhf}; the phantom energy can disrupt all
gravitationally bound structures i.e from galaxies to black holes
\cite{babi,Ness,mota1,mota2,babi2,babi3}; it can produce infinite
expansion of the Universe in a finite time thus causing the `big
rip' \cite{cald}.

From above equations, we get
\begin{equation}
\dot{\phi}^{2}=-\frac{(1+w_{m})}{\epsilon}\rho_{m}+\frac{1}{4\pi\epsilon
G}\left[-\dot{H}+\frac{(\xi-1)}{T_1}H+\frac{k}{a^{2}}  \right],
\end{equation}
and
\begin{equation}
V=\frac{(w_{m}-1)}{2}\rho_{m}+\frac{1}{8\pi
G}\left[\dot{H}+3H^{2}+\frac{5(\xi-1)}{T_1}H+\frac{2k}{a^{2}}.
\right]
\end{equation}

$\bullet{}$ For emergent scenario, we get the expressions for
$\phi$ and $V$ as
\begin{equation}
\phi=\int
\sqrt{-\frac{(1+w_{m})\rho_{0}a_{0}^{-3(1+w_{m}+\delta)}}{\epsilon\left(\lambda+e^{\mu
T_1}\right)^{3n(1+w_{m}+\delta)}}
 +\frac{1}{4\pi\epsilon G} \left\{ -\frac{n\lambda\mu^{2}
e^{\mu T_1}}{ \left(\lambda+e^{\mu
T_1}\right)^{2}}+\frac{(\xi-1)n\mu e^{\mu T_1}}{T_1(\lambda+e^{\mu
T_1})} +\frac{k~a_{0}^{-2}}{ \left(\lambda+e^{\mu
T_1}\right)^{2n}} \right\} } ~~~ dT_1,
\end{equation}
and
\begin{equation}
V=\frac{(w_{m}-1)\rho_{0}a_{0}^{-3(1+w_{m}+\delta)}}{2\left(\lambda+e^{\mu
T_1}\right)^{3n(1+w_{m}+\delta)}}
 +\frac{1}{8\pi G} \left\{ \frac{n\mu^{2}e^{\mu
T_1}(\lambda+3ne^{\mu T_1})}{ \left(\lambda+e^{\mu
T_1}\right)^{2}}+\frac{5(\xi-1)n\mu e^{\mu T_1}}{T_1(\lambda+e^{\mu
T_1})} +\frac{2k~a_{0}^{-2}}{ \left(\lambda+e^{\mu T_1}\right)^{2n}}
\right\}.
\end{equation}

\begin{figure}
\includegraphics[height=2.0in]{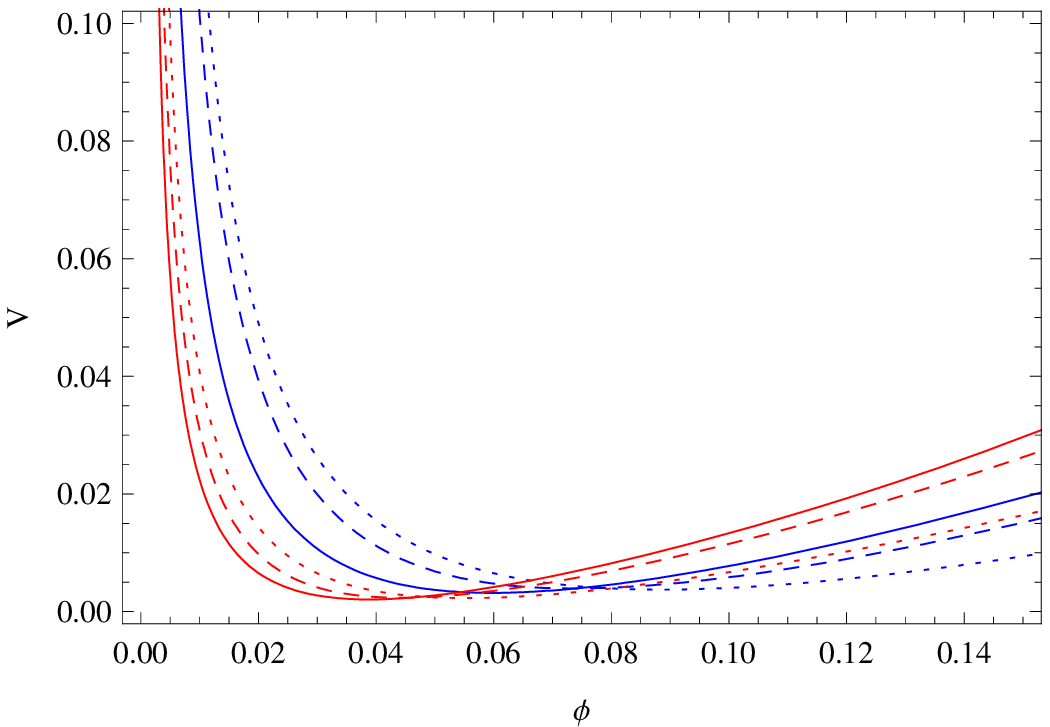}~~~~~~~~
\includegraphics[height=2.0in]{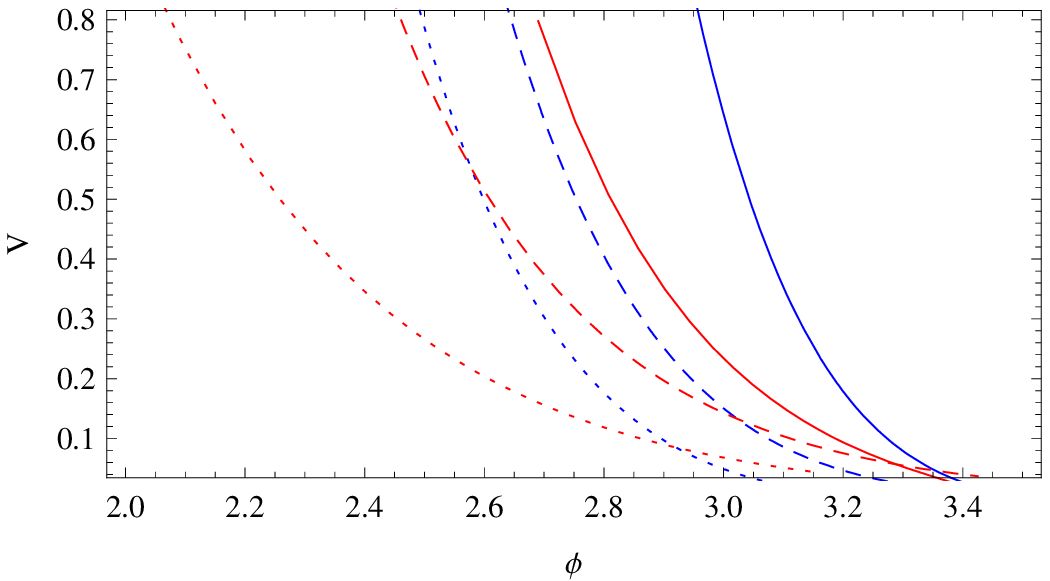}\\
\vspace{2mm}
~~~~Fig.1~~~~~~~~~~~~~~~~~~~~~~~~~~~~~~~~~~~~~~~~~~~~~~~~~~~~~~~~~~~~~~Fig.2~~\\
\vspace{6mm}
\includegraphics[height=2.0in]{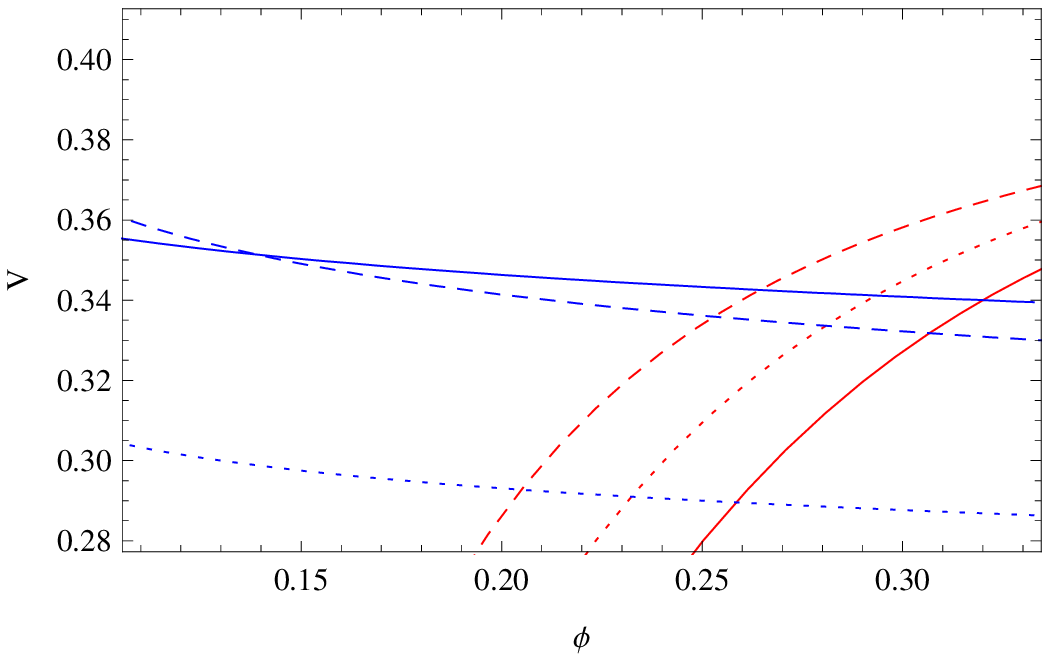}\\
\vspace{2mm} Fig.3

Figs.1-3 show the variations of $V$ against quintessence or
phantom field $\phi$ in the emergent, logamediate and intermediate
scenarios respectively. Solid, dash and dotted lines represent
$k=-1,+1,0$ respectively. Blue and red lines represent
quintessence field ($\epsilon=+1$) and phantom field
($\epsilon=-1$) respectively.

\end{figure}

$\bullet{}$ For logamediate scenario, we get the expressions for
$\phi$ and $V$ as
\begin{equation}
\phi=\int
\sqrt{-\frac{(1+w_{m})\rho_{0}}{\epsilon}~e^{-3A(1+w_{m}+\delta)(\ln
T_1)^{\alpha}}
 +\frac{1}{4\pi\epsilon G} \left\{ \frac{A\alpha}{T_1^{2}}(\ln
T_1)^{\alpha-2}(1-\alpha+\xi \ln T_1) +k~e^{-2A(\ln T_1)^{\alpha}}
\right\} } ~~~ dT_1
\end{equation}
and
\begin{equation}
V=\frac{(w_{m}-1)\rho_{0}}{2}~e^{-3A(1+w_{m}+\delta)(\ln
T_1)^{\alpha}}
 +\frac{1}{8\pi G} \left[ \frac{A\alpha}{T_1^{2}}(\ln
T_1)^{\alpha-2}\{\alpha-1+(5\xi-6)\ln T_1 +3A\alpha(\ln
T_1)^{\alpha}\} +2k~e^{-2A(\ln T_1)^{\alpha}} \right].
\end{equation}

$\bullet{}$ For intermediate scenario, we get the expressions for
$\phi$ and $V$ as
\begin{equation}
\phi=\int
\sqrt{-\frac{(1+w_{m})\rho_{0}}{\epsilon}~e^{-3B(1+w_{m}+\delta)T_1^{\beta}}
 +\frac{1}{4\pi\epsilon G} \left\{ B\beta(\xi-\beta)T_1^{\beta-2} +k~e^{-2B T_1^{\beta}}
\right\} } ~~~ dT_1,
\end{equation}
and
\begin{equation}
V=\frac{(w_{m}-1)\rho_{0}}{2}~e^{-3B(1+w_{m}+\delta) T_1^{\beta}}
 +\frac{1}{8\pi G} \left[B\beta T_1^{\beta-2}(5\xi+\beta+3B\beta T_1^{\beta})
+2k~e^{-2B T_1^{\beta}} \right].
\end{equation}

In figures 1, 2 and 3, we have plotted the potentials against the
scalar fields for the quintessence and phantom fields in emergent,
logamediate and intermediate scenarios of the universe
respectively in fractional action cosmology. It has been observed
in figure 1 that after gradual decay, the potential starts
increasing with scalar field for quintessence as well as phantom
field models of dark energy in the emergent scenario of the
universe irrespective of its type of curvature. On the contrary,
when logamediate scenario is considered, the figure 2 exhibits a
continuous decay in the potential $V$ with increase in the scalar
field $\phi$. A different behavior is observed in figure 3 that
depicts the behavior of the potential $V$ against scalar field
$\phi$ in the case of intermediate scenario of the universe. The
blue lines in this figure show a continuous decay in $V$ with
increase in $\phi$ for quintessence model. However, the red lines
exhibit an increasing pattern of $V$ with scalar field $\phi$.

\subsection{Tachyonic field}

A rolling tachyon has an interesting equation of state whose state
parameter smoothly interpolates between $-1$ and $0$ \cite{Gib1}.
Thus, tachyon can be realized as a suitable candidate for the
inflation at high energy \cite{Maz1} as well as a source of dark
energy depending on the form of the tachyon potential \cite{Padm}.
Therefore it becomes meaningful to reconstruct tachyon potential
$V(\phi)$ from some dark energy models. An action for tachyon scalar
$\phi$ is given by Born-Infeld like action

\begin{eqnarray}
S = - \int {d^4x} \sqrt{-g} V(\phi) \sqrt{1 - g^{ij} \partial_i
\phi \partial_j \phi}
\end{eqnarray}
where $V(\phi)$ is the tachyon potential. Energy-momentum tensor
components for tachyon scalar $\phi$ are obtained as

\begin{eqnarray}
T_{ij} = V(\phi) \left[ \frac{\partial_i \phi \partial_j
\phi}{\sqrt{1 - g^{ij}\partial_i \phi \partial_j \phi}} + g_{ij}
\sqrt{1 - g^{kl}\partial_k \phi \partial_l \phi}\right]
\end{eqnarray}

The energy density $\rho_{\phi}$ pressure $p_{\phi}$ due to the
tachyonic field $\phi$ have the expressions
\begin{eqnarray}
\rho_{\phi}&=&\frac{V(\phi)}{\sqrt{1-\epsilon{\dot{\phi}}^{2}}},\\
p_{\phi}&=&-V(\phi) \sqrt{1-\epsilon{\dot{\phi}}^{2}},
\end{eqnarray}
where $V(\phi)$ is the relevant potential for the tachyonic field
$\phi$. It is to be seen that $\frac{p_{\phi}}
{\rho_{\phi}}=-1+\epsilon\dot{\phi}^{2}$ $>-1$ or $<-1$
accordingly as normal tachyon ($\epsilon=+1$) or phantom tachyon
($\epsilon=-1$).\\

From above, we get

\begin{eqnarray*}
\dot{\phi}^{2}=\left[-\frac{(1+w_{m})}{\epsilon}\rho_{m}+\frac{1}{4\pi\epsilon
G}\left\{-\dot{H}+\frac{(\xi-1)}{T_1}H+\frac{k}{a^{2}} \right\}
\right]
\end{eqnarray*}
\begin{equation}
\times \left[-\rho_{m}+\frac{3}{8\pi
G}\left\{H^{2}+\frac{2(\xi-1)}{T_1}H+\frac{k}{a^{2}} \right\}
\right]^{-1}
\end{equation}
and
\begin{eqnarray*}
V=\left[w_{m}\rho_{m}+\frac{1}{8\pi
G}\left\{2\dot{H}+3H^{2}+\frac{4(\xi-1)}{T_1}H+\frac{k}{a^{2}}
\right\} \right]^{\frac{1}{2}}
\end{eqnarray*}
\begin{equation}
\times \left[-\rho_{m}+\frac{3}{8\pi
G}\left\{H^{2}+\frac{2(\xi-1)}{T_1}H+\frac{k}{a^{2}} \right\}
\right]^{\frac{1}{2}}
\end{equation}

$\bullet{}$ For emergent scenario, we get the expressions for
$\phi$ and $V$ as
\begin{eqnarray*}
\phi= \int
\left[-\frac{(1+w_{m})\rho_{0}a_{0}^{-3(1+w_{m}+\delta)}}{\epsilon\left(\lambda+e^{\mu
T_1}\right)^{3n(1+w_{m}+\delta)}}
 +\frac{1}{4\pi\epsilon G} \left\{ -\frac{n\lambda\mu^{2}
e^{\mu T_1}}{ \left(\lambda+e^{\mu
T_1}\right)^{2}}+\frac{(\xi-1)n\mu e^{\mu T_1}}{T_1(\lambda+e^{\mu
T_1})} +\frac{k~a_{0}^{-2}}{ \left(\lambda+e^{\mu T_1}\right)^{2n}}
\right\} \right]^{\frac{1}{2}}
\end{eqnarray*}
\begin{equation}
\times
\left[-\frac{\rho_{0}a_{0}^{-3(1+w_{m}+\delta)}}{\left(\lambda+e^{\mu
T_1}\right)^{3n(1+w_{m}+\delta)}}
 +\frac{3}{8\pi G} \left\{ \frac{n^{2}\mu^{2}
e^{2\mu T_1}}{ \left(\lambda+e^{\mu
T_1}\right)^{2}}+\frac{2(\xi-1)n\mu e^{\mu t}}{T_1(\lambda+e^{\mu
T_1})} +\frac{k~a_{0}^{-2}}{ \left(\lambda+e^{\mu T_1}\right)^{2n}}
\right\} \right]^{-\frac{1}{2}} ~~dT_1
\end{equation}
and
\begin{eqnarray*}
V= \left[\frac{w_{m
}\rho_{0}a_{0}^{-3(1+w_{m}+\delta)}}{\left(\lambda+e^{\mu
T_1}\right)^{3n(1+w_{m}+\delta)}}
 +\frac{1}{8\pi G} \left\{ \frac{n\mu^{2}e^{\mu
T_1}(2\lambda+3ne^{\mu T_1})}{ \left(\lambda+e^{\mu
T_1}\right)^{2}}+\frac{4(\xi-1)n\mu e^{\mu T_1}}{T_1(\lambda+e^{\mu
T_1})} +\frac{k~a_{0}^{-2}}{ \left(\lambda+e^{\mu T_1}\right)^{2n}}
\right\} \right]^{\frac{1}{2}}
\end{eqnarray*}
\begin{equation}
\times
\left[-\frac{\rho_{0}a_{0}^{-3(1+w_{m}+\delta)}}{\left(\lambda+e^{\mu
T_1}\right)^{3n(1+w_{m}+\delta)}}
 +\frac{3}{8\pi G} \left\{ \frac{n^{2}\mu^{2}
e^{2\mu T_1}}{ \left(\lambda+e^{\mu
T_1}\right)^{2}}+\frac{2(\xi-1)n\mu e^{\mu T_1}}{T_1(\lambda+e^{\mu
T_1})} +\frac{k~a_{0}^{-2}}{ \left(\lambda+e^{\mu T_1}\right)^{2n}}
\right\} \right]^{\frac{1}{2}}.
\end{equation}

\begin{figure}
\includegraphics[height=2.7in]{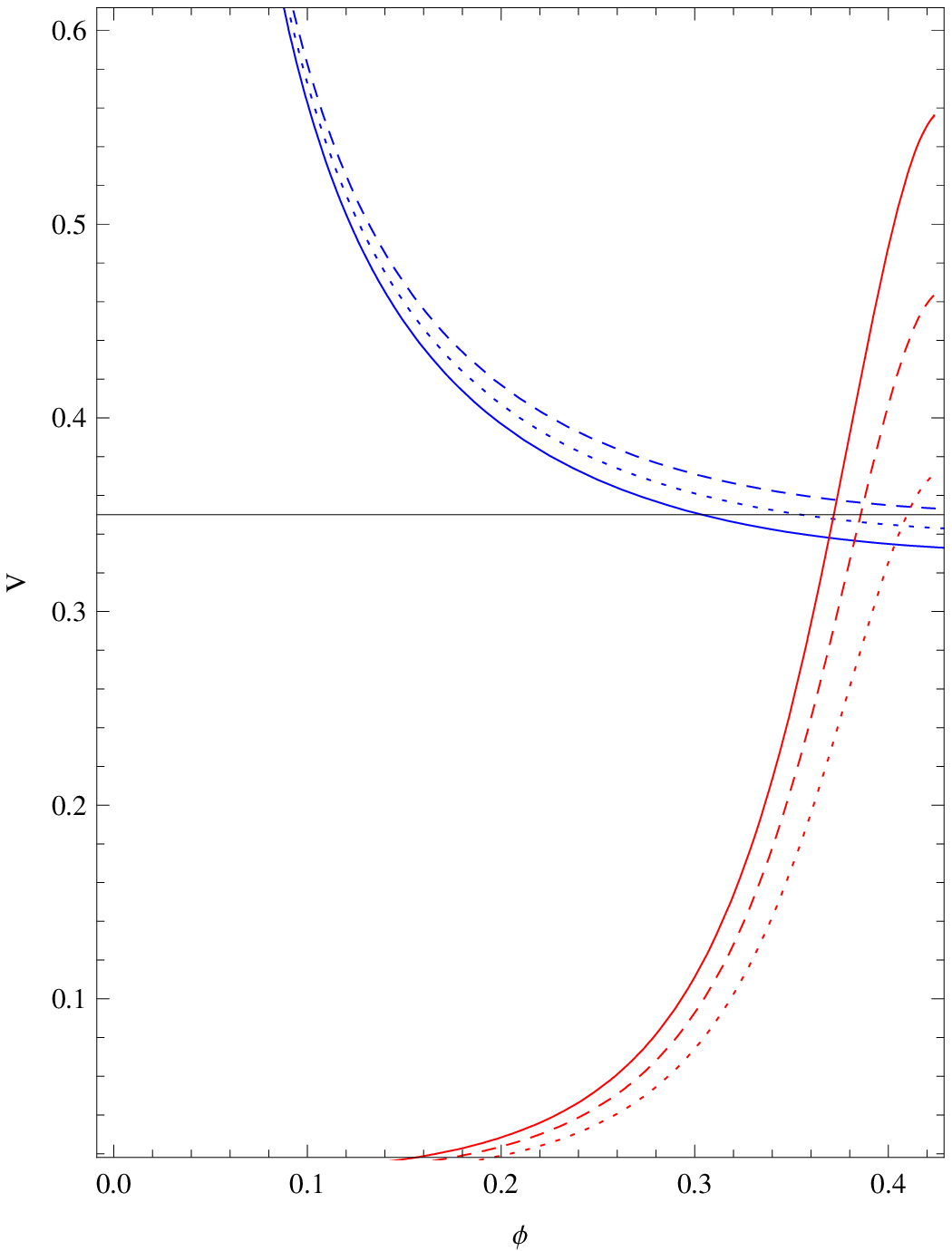}~~~~~~~~
\includegraphics[height=2.0in]{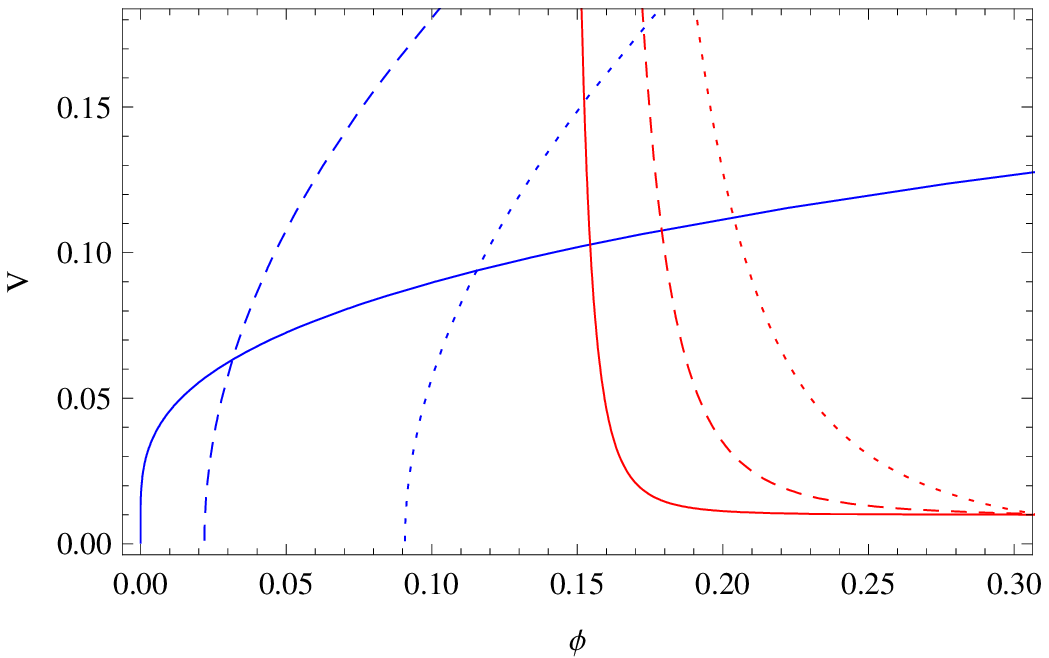}\\
\vspace{2mm}
~~~~Fig.4~~~~~~~~~~~~~~~~~~~~~~~~~~~~~~~~~~~~~~~~~~~~~~~~~~~~~~~~~~~~~~Fig.5~~\\
\vspace{6mm}
\includegraphics[height=3.0in]{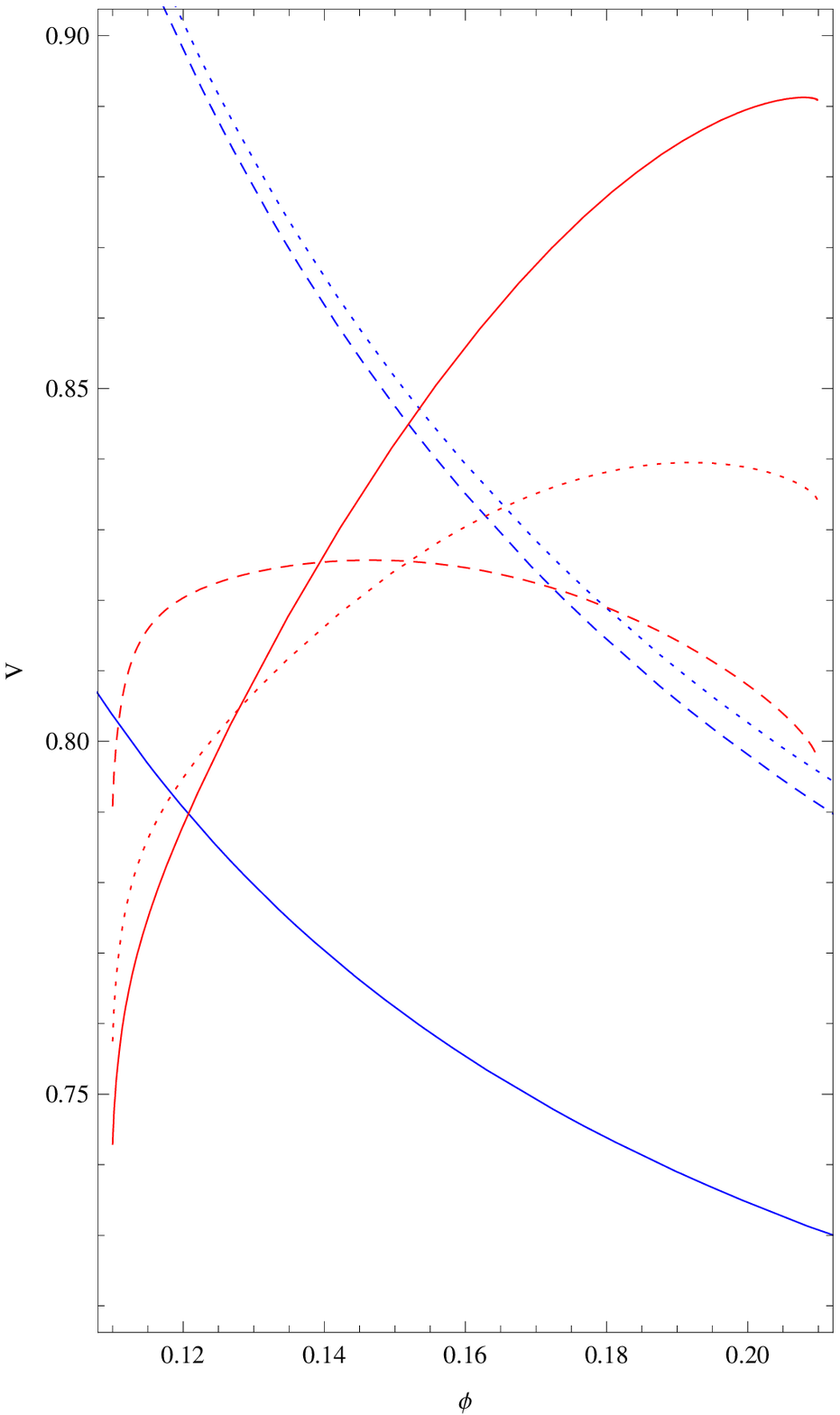}\\
\vspace{2mm} Fig.6

Figs.4-6 show the variations of $V$ against tachyonic field $\phi$
in the emergent, logamediate and intermediate scenarios
respectively. Solid, dash and dotted lines represent $k=-1,+1,0$
respectively. Blue and red lines represent normal tachyonic field
($\epsilon=+1$) and phantom tachyonic field ($\epsilon=-1$)
respectively.
\end{figure}

$\bullet{}$ For logamediate scenario, we get the expressions for
$\phi$ and $V$ as
\begin{eqnarray*}
\phi= \int
\left[-\frac{(1+w_{m})\rho_{0}}{\epsilon}~e^{-3A(1+w_{m}+\delta)(\ln
T_1)^{\alpha}}
 +\frac{1}{4\pi\epsilon G} \left\{ \frac{A\alpha}{T_1^{2}}(\ln
T_1)^{\alpha-2}(1-\alpha+\xi \ln T_1) +k~e^{-2A(\ln T_1)^{\alpha}}
\right\} \right]^{\frac{1}{2}}
\end{eqnarray*}
\begin{equation}
\times \left[-\rho_{0}~e^{-3A(1+w_{m}+\delta)(\ln T_1)^{\alpha}}
 +\frac{3}{8\pi G} \left\{ \frac{A\alpha}{T_1^{2}}(\ln
T_1)^{\alpha-1}\{ A\alpha(\ln T_1)^{\alpha-1}+2(\xi-1)  \}
+k~e^{-2A(\ln T_1)^{\alpha}} \right\} \right]^{-\frac{1}{2}} ~~dT_1
\end{equation}
and
\begin{eqnarray*}
V=\left[-\rho_{0}~e^{-3A(1+w_{m}+\delta)(\ln T_1)^{\alpha}}
 +\frac{3}{8\pi G} \left\{ \frac{A\alpha}{T_1^{2}}(\ln
T_1)^{\alpha-1}\{ A\alpha(\ln T_1)^{\alpha-1}+2(\xi-1)  \}
+k~e^{-2A(\ln T_1)^{\alpha}} \right\}\right]^{\frac{1}{2}}
\end{eqnarray*}
\begin{equation}
\times \left[w_{m }\rho_{0}~e^{-3A(1+w_{m}+\delta)(\ln
T_1)^{\alpha}}
 +\frac{1}{8\pi G} \left\{ \frac{A\alpha}{t^{2}}(\ln
T_1)^{\alpha-2}\{2(\alpha-1)+2(\xi-3)\ln t +3A\alpha(\ln
T_1)^{\alpha} \} +k~e^{-2A(\ln T_1)^{\alpha}} \right\}
\right]^{\frac{1}{2}}
\end{equation}

$\bullet{}$ For intermediate scenario, we get the expressions for
$\phi$ and $V$ as
\begin{eqnarray*}
\phi= \int
\left[-\frac{(1+w_{m})\rho_{0}}{\epsilon}~e^{-3B(1+w_{m}+\delta)T_1^{\beta}}
 +\frac{1}{4\pi\epsilon G} \left\{ B\beta(\xi-\beta)T_1^{\beta-2} +k~e^{-2B T_1^{\beta}}
\right\} \right]^{\frac{1}{2}}
\end{eqnarray*}
\begin{equation}
\times \left[-\rho_{0}~e^{-3B(1+w_{m}+\delta)T_1^{\beta}}
 +\frac{3}{8\pi G} \left\{ B\beta T_1^{\beta-2} (2(\xi-1)+B\beta T_1^{\beta}) +k~e^{-2B T_1^{\beta}}
\right\} \right]^{-\frac{1}{2}} ~~dT_1
\end{equation}
and
\begin{eqnarray*}
V=\left[-\rho_{0}~e^{-3B(1+w_{m}+\delta)T_1^{\beta}}
 +\frac{3}{8\pi G} \left\{ B\beta T_1^{\beta-2} (2(\xi-1)+B\beta T_1^{\beta}) +k~e^{-2B T_1^{\beta}}
\right\} \right]^{\frac{1}{2}}
\end{eqnarray*}
\begin{equation}
\times \left[w_{m}\rho_{0}~e^{-3B(1+w_{m}+\delta)T_1^{\beta}}
 +\frac{1}{8\pi G} \left\{ B\beta T_1^{\beta-2} (2(2\xi+\beta-3)+3B\beta T_1^{\beta}) +k~e^{-2B T_1^{\beta}}
\right\} \right]^{\frac{1}{2}}.
\end{equation}

In figure 4, the $V$-$\phi$ plot for normal tachyon and phantom
tachyon models of dark energy is presented for emergent scenario
of the universe. Potential of normal tachyon exhibits decaying
pattern.  However, it shows increasing pattern for phantom
tachyonic field $\phi$. It happens irrespective of the curvature
of the universe. In the logamediate scenario (figure 5) the
potentials for normal tachyon and phantom tachyon exhibit
increasing and decreasing behavior respectively with increase in
the scalar field $\phi$. From figure 6 we see a continuous decay
in the potential for normal tachyonic field in the intermediate
scenario. However, in this scenario, the behavior of the potential
varies with the curvature of the universe characterized by
interacting phantom tachyonic field. For $k=-1,~1$, the potential
increases with phantom tachyonic field and for $k=0$, it decays
after increasing initially.

\subsection{k-essence}

In the kinetically driven scalar field theory, we have {\em
non-canonical} kinetic energy term with {\em no} potential. Scalars,
modelling this theory, are popularly known as {\em k-essence}.
Motivated by Born-Infeld action of String Theory, it was used as a
source to explain the mechanism for producing the late time
acceleration of the universe. This model is given by the action
\cite{k}
\begin{equation}
S = \int {d^4x} \sqrt{-g} {\tilde {\cal L}}({\tilde \phi}, {\tilde
X}),
\end{equation}
with
\begin{equation}
{\tilde {\cal L}}({\tilde \phi}, {\tilde X}) = K({\tilde
\phi}){\tilde X} + L({\tilde \phi}){\tilde X}^2,
\end{equation}
ignoring higher order terms of
\begin{equation}
{\tilde X} = \frac{1}{2} g^{ij}\partial_i {\tilde \phi} \partial_j
{\tilde \phi}.
\end{equation}
Using the following transformations, $\phi = \int {d {\tilde \phi}}
\sqrt{|L({\tilde \phi})|/K({\tilde \phi})}, $ $ X = \frac{|L|}{K}
{\tilde X}$ and $ V(\phi) = K^2/|L|$~, the action can be rewritten
as
\begin{equation}
S = \int {d^4x} \sqrt{-g} V(\phi) {\cal L}(X),
\end{equation}
with
\begin{equation}
{\cal L}(X) = X - X^2.
\end{equation}

From the action, the energy-momentum tensor components can be
written as
\begin{equation}
T_{ij} = V(\phi) \Big[\frac{d {\cal L}}{dX}\partial_i \phi
\partial_j \phi - g_{ij} {\cal L}\Big].
\end{equation}
The energy density and pressure of k-essence scalar field $\phi$ are
given by
\begin{equation}
\rho_{k}=V(\phi)(-X+3X^{2}),
\end{equation}
and
\begin{equation}
p_{k}=V(\phi)(-X+X^{2}),
\end{equation}
where $\phi$ is the scalar field having kinetic energy $X=
\frac{1}{2}\dot{\phi}^{2}$ and $V(\phi)$ is
the k-essence potential.\\

From above, we get

\begin{eqnarray*}
\dot{\phi}^{2}=\left[2(w_{m}-1)\rho_{m}+\frac{1}{2\pi
G}\left\{\dot{H}+3H^{2}+\frac{5(\xi-1)}{T_1}H+\frac{2k}{a^{2}}
\right\} \right]
\end{eqnarray*}
\begin{equation}
\times \left[(3w_{m}-1)\rho_{m}+\frac{3}{4\pi
G}\left\{\dot{H}+2H^{2}+\frac{3(\xi-1)}{T_1}H+\frac{k}{a^{2}}
\right\} \right]^{-1},
\end{equation}
and
\begin{eqnarray*}
V=\left[(3w_{m}-1)\rho_{m}+\frac{3}{4\pi
G}\left\{\dot{H}+2H^{2}+\frac{3(\xi-1)}{T_1}H+\frac{k}{a^{2}}
\right\} \right]^{2}
\end{eqnarray*}
\begin{equation}
\times  \left[2(w_{m}-1)\rho_{m}+\frac{1}{2\pi
G}\left\{\dot{H}+3H^{2}+\frac{5(\xi-1)}{T_1}H+\frac{2k}{a^{2}}
\right\} \right]^{-1}.
\end{equation}

$\bullet{}$ For emergent scenario, we have

\begin{eqnarray*}
\phi=\int
\left[\frac{2(w_{m}-1)\rho_{0}a_{0}^{-3(1+w_{m}+\delta)}}{\left(\lambda+e^{\mu
T_1}\right)^{3n(1+w_{m}+\delta)}}
 +\frac{1}{2\pi G} \left\{ \frac{n\mu^{2}e^{\mu
T_1}(\lambda+3ne^{\mu T_1})}{ \left(\lambda+e^{\mu
T_1}\right)^{2}}+\frac{5(\xi-1)n\mu e^{\mu T_1}}{T_1(\lambda+e^{\mu
T_1})} +\frac{2k~a_{0}^{-2}}{ \left(\lambda+e^{\mu T_1}\right)^{2n}}
\right\} \right]^{\frac{1}{2}}
\end{eqnarray*}
\begin{equation}
\times
\left[\frac{(3w_{m}-1)\rho_{0}a_{0}^{-3(1+w_{m}+\delta)}}{\left(\lambda+e^{\mu
T_1}\right)^{3n(1+w_{m}+\delta)}}
 +\frac{3}{4\pi G} \left\{ \frac{n\mu^{2}e^{\mu
T_1}(\lambda+2ne^{\mu T_1})}{ \left(\lambda+e^{\mu
T_1}\right)^{2}}+\frac{3(\xi-1)n\mu e^{\mu T_1}}{T_1(\lambda+e^{\mu
T_1})} +\frac{k~a_{0}^{-2}}{ \left(\lambda+e^{\mu T_1}\right)^{2n}}
\right\} \right]^{-\frac{1}{2}}~dt.
\end{equation}
and
\begin{eqnarray*}
V=\left[\frac{(3w_{m}-1)\rho_{0}a_{0}^{-3(1+w_{m}+\delta)}}{\left(\lambda+e^{\mu
T_1}\right)^{3n(1+w_{m}+\delta)}}
 +\frac{3}{4\pi G} \left\{ \frac{n\mu^{2}e^{\mu
T_1}(\lambda+2ne^{\mu T_1})}{ \left(\lambda+e^{\mu
T_1}\right)^{2}}+\frac{3(\xi-1)n\mu e^{\mu T_1}}{T_1(\lambda+e^{\mu
T_1})} +\frac{k~a_{0}^{-2}}{ \left(\lambda+e^{\mu T_1}\right)^{2n}}
\right\} \right]^{2}
\end{eqnarray*}
\begin{equation}
\times
\left[\frac{2(w_{m}-1)\rho_{0}a_{0}^{-3(1+w_{m}+\delta)}}{\left(\lambda+e^{\mu
T_1}\right)^{3n(1+w_{m}+\delta)}}
 +\frac{1}{2\pi G} \left\{ \frac{n\mu^{2}e^{\mu
T_1}(\lambda+3ne^{\mu T_1})}{ \left(\lambda+e^{\mu
T_1}\right)^{2}}+\frac{5(\xi-1)n\mu e^{\mu T_1}}{T_1(\lambda+e^{\mu
T_1})} +\frac{2k~a_{0}^{-2}}{ \left(\lambda+e^{\mu T_1}\right)^{2n}}
\right\} \right]^{-1}.
\end{equation}

$\bullet{}$ For logamediate scenario, we get the expressions for
$\phi$ and $V$ as

\begin{eqnarray*}
\phi= \int \left[2(w_{m}-1)\rho_{0}~e^{-3A(1+w_{m}+\delta)(\ln
T_1)^{\alpha}}
 +\frac{1}{2\pi G} \left\{ \frac{A\alpha}{T_1^{2}}(\ln
T_1)^{\alpha-2}(\alpha-1+(5\xi-6)\ln T_1+3A\alpha(\ln T_1)^{\alpha})
+2k~e^{-2A(\ln T_1)^{\alpha}} \right\} \right]^{\frac{1}{2}}
\end{eqnarray*}
\begin{equation}
\times \left[(3w_{m}-1)\rho_{0}~e^{-3A(1+w_{m}+\delta)(\ln
T_1)^{\alpha}}
 +\frac{3}{4\pi G} \left\{ \frac{A\alpha}{T_1^{2}}(\ln
T_1)^{\alpha-2}(\alpha-1+(3\xi-4)\ln T_1+2A\alpha(\ln T_1)^{\alpha})
+k~e^{-2A(\ln T_1)^{\alpha}} \right\} \right]^{-\frac{1}{2}} ~~dT_1
\end{equation}
and
\begin{eqnarray*}
V=\left[(3w_{m}-1)\rho_{0}~e^{-3A(1+w_{m}+\delta)(\ln T_1)^{\alpha}}
 +\frac{3}{4\pi G} \left\{ \frac{A\alpha}{T_1^{2}}(\ln
T_1)^{\alpha-2}(\alpha-1+(3\xi-4)\ln T_1+2A\alpha(\ln T_1)^{\alpha})
+k~e^{-2A(\ln T_1)^{\alpha}} \right\} \right]^{2}
\end{eqnarray*}
\begin{equation}
\times \left[2(w_{m}-1)\rho_{0}~e^{-3A(1+w_{m}+\delta)(\ln
T_1)^{\alpha}}
 +\frac{1}{2\pi G} \left\{ \frac{A\alpha}{T_1^{2}}(\ln
T_1)^{\alpha-2}(\alpha-1+(5\xi-6)\ln T_1+3A\alpha(\ln T_1)^{\alpha})
+2k~e^{-2A(\ln T_1)^{\alpha}} \right\} \right]^{-1}
\end{equation}

$\bullet{}$ For intermediate scenario, we get the expressions for
$\phi$ and $V$ as

\begin{eqnarray*}
\phi= \int
\left[2(w_{m}-1)\rho_{0}~e^{-3B(1+w_{m}+\delta)T_1^{\beta}}
 +\frac{1}{2\pi G} \left\{ B\beta(5\xi+\beta-6+3B\beta T_1^{\beta})T_1^{\beta-2} +2k~e^{-2B T_1^{\beta}}
\right\} \right]^{\frac{1}{2}}
\end{eqnarray*}
\begin{equation}
\times \left[(3w_{m}-1)\rho_{0}~e^{-3B(1+w_{m}+\delta)T_1^{\beta}}
 +\frac{3}{4\pi G} \left\{ B\beta(3\xi+\beta-4+2B\beta T_1^{\beta})T_1^{\beta-2} +k~e^{-2B T_1^{\beta}}
\right\} \right]^{-\frac{1}{2}} ~~dT_1,
\end{equation}
and
\begin{eqnarray*}
V=\left[(3w_{m}-1)\rho_{0}~e^{-3B(1+w_{m}+\delta)T_1^{\beta}}
 +\frac{3}{4\pi G} \left\{ B\beta(3\xi+\beta-4+2B\beta T_1^{\beta})T_1^{\beta-2} +k~e^{-2B T_1^{\beta}}
\right\} \right]^{2}
\end{eqnarray*}
\begin{equation}
\times \left[2(w_{m}-1)\rho_{0}~e^{-3B(1+w_{m}+\delta)T_1^{\beta}}
 +\frac{1}{2\pi G} \left\{ B\beta(5\xi+\beta-6+3B\beta T_1^{\beta})T_1^{\beta-2}
 +2k~e^{-2B T_1^{\beta}}
\right\} \right]^{-1}.
\end{equation}

From figures 7, 8 and 9 we see that for interacting k-essence the
potential $V$ always decreases with increase in the scalar field
$\phi$ in all of the three scenarios and it happens for open,
closed and flat universes.

\begin{figure}
\includegraphics[height=3.0in]{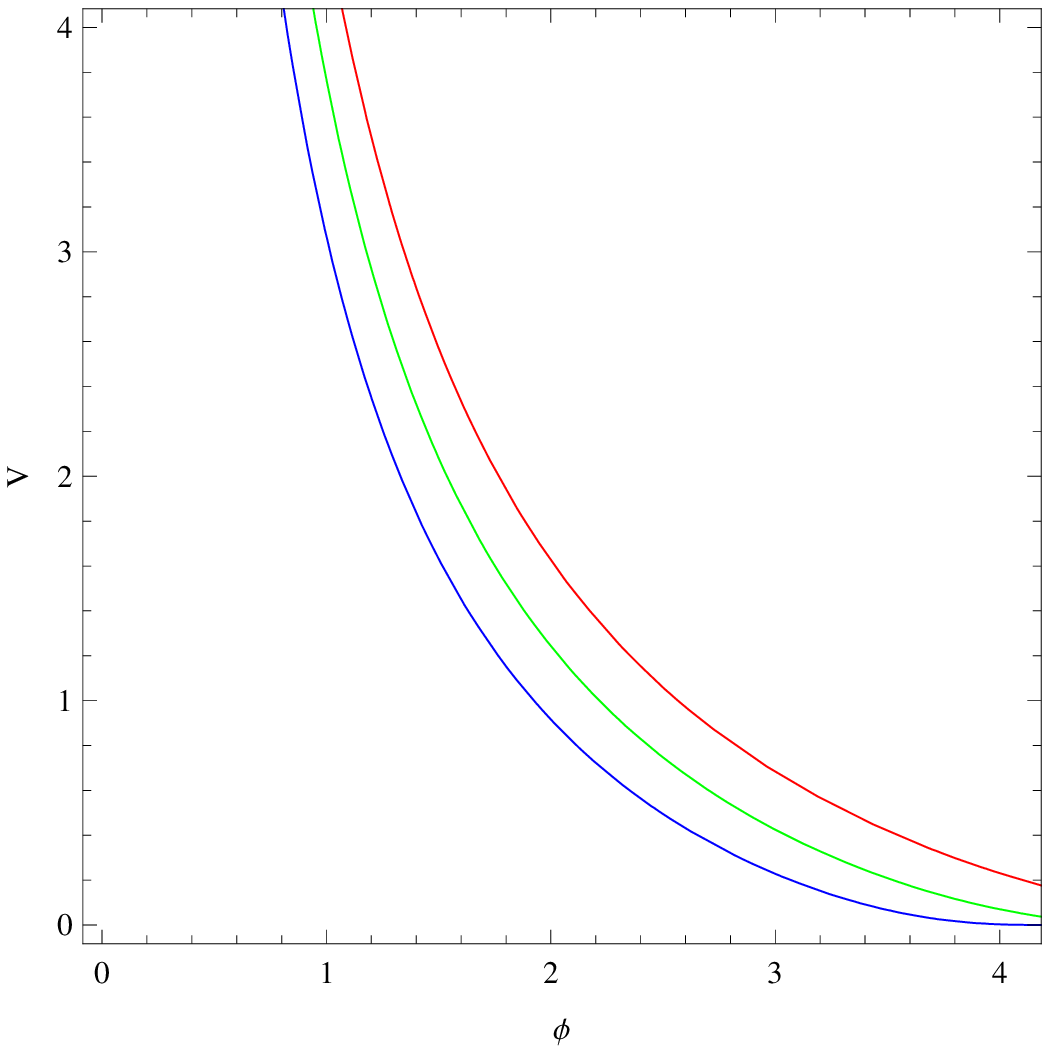}~~~~~~~~
\includegraphics[height=2.7in]{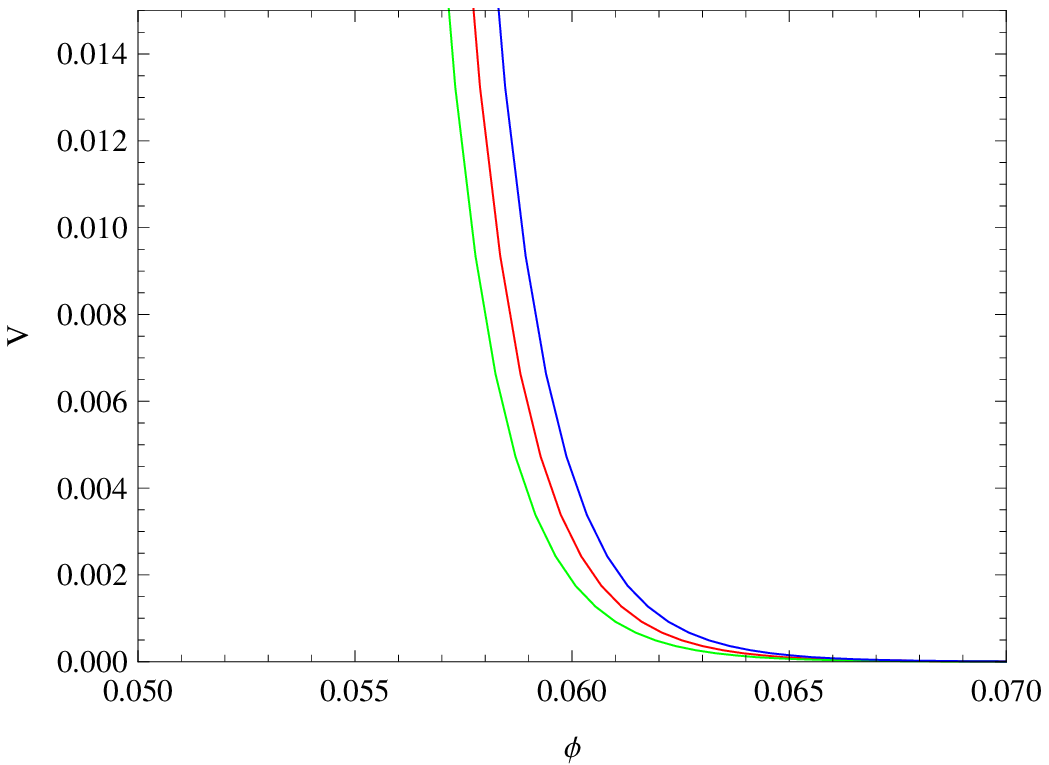}\\
\vspace{2mm}
~~~~Fig.7~~~~~~~~~~~~~~~~~~~~~~~~~~~~~~~~~~~~~~~~~~~~~~~~~~~~~~~~~~~~~~Fig.8~~\\
\vspace{6mm}
\includegraphics[height=3.0in]{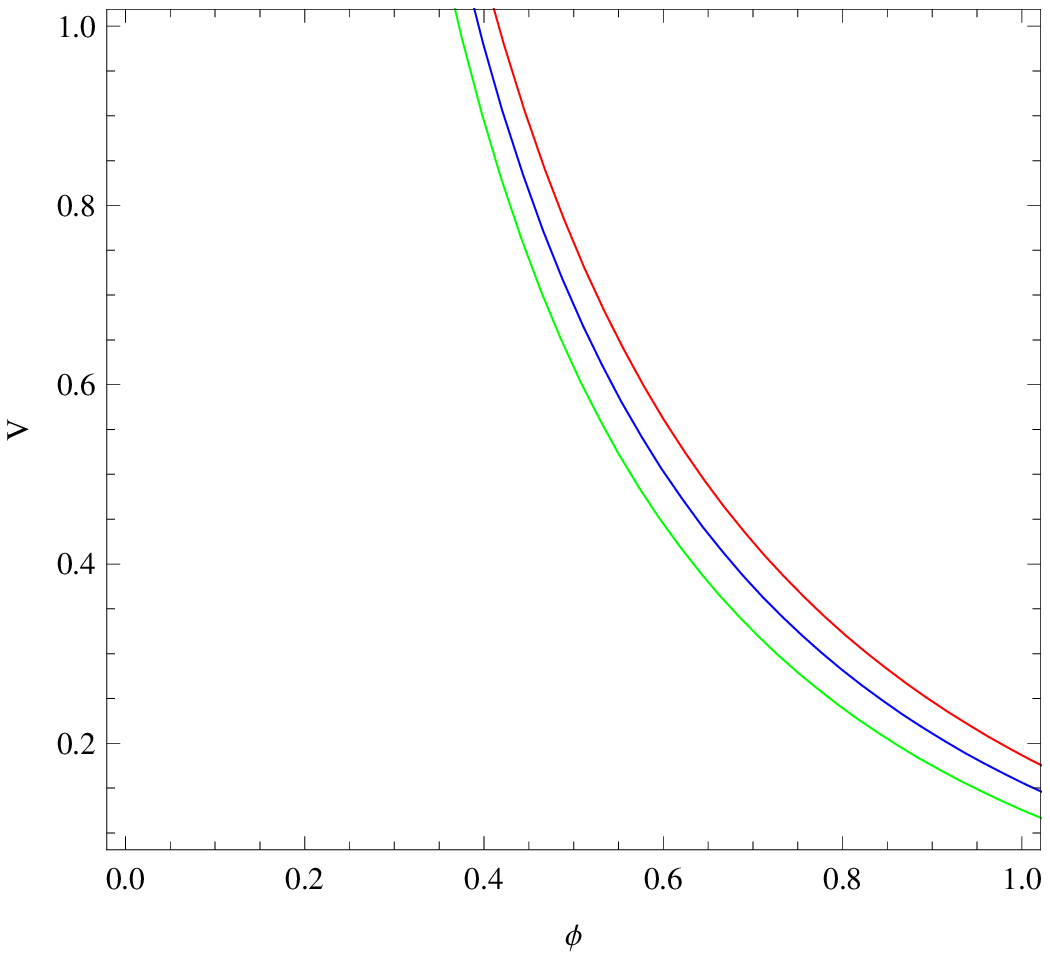}\\
\vspace{2mm} Fig.9

Figs.7-9 show the variations of $V$ against k-essence field $\phi$
in the emergent, logamediate and intermediate scenarios
respectively. Red, green and blue lines represent $k=-1,+1,0$
respectively.
\end{figure}

\subsection{DBI-essence}

Consider that the dark energy scalar field is a Dirac-Born-Infeld
(DBI) scalar field. In this case, the action of the field be written
as \cite{dbi}
\begin{equation}
S_{D}=-\int d^4x
a^3(t)\left[T(\phi)\sqrt{1-\frac{\dot{\phi^{2}}}{T(\phi)}}+V(\phi)-T(\phi)\right],
\end{equation}
where $T(\phi)$ is the warped brane tension and $V(\phi)$ is the DBI
potential. The energy density and pressure of the DBI-essence scalar
field are respectively given by
\begin{equation}
\rho_{D}=(\gamma-1)T(\phi)+V(\phi),
\end{equation}
and
\begin{equation}
p_{D}=\frac{\gamma-1}{\gamma}T(\phi)-V(\phi),
\end{equation}
where $\gamma$ is given by
\begin{equation}
\gamma=\frac{1}{\sqrt{1-\frac{\dot{\phi}^{2}}{T(\phi)}}}.
\end{equation}

Now we consider here particular case $\gamma =$ constant. In this
case, for simplicity, we assume $T(\phi)= T_{0} \dot{\phi}^{2}$
$(T_{0}
> 1)$. So we have $\gamma = \sqrt{\frac{T_{0}}{T_{0}-1}}$. In this case the
expressions for $\phi$, $T(\phi)$ and $V(\phi)$ are given by

\begin{equation}
\dot{\phi}^{2}=\sqrt{\frac{T_{0}-1}{T_{0}}}\left[-(1+w_{m})\rho_{m}+\frac{1}{4\pi
G}\left(-\dot{H}+\frac{\xi-1}{T_1}H+\frac{k}{a^{2}}  \right)
\right].
\end{equation}
\begin{equation}
T=\sqrt{T_{0}(T_{0}-1)}\left[-(1+w_{m})\rho_{m}+\frac{1}{4\pi
G}\left(-\dot{H}+\frac{\xi-1}{t}H+\frac{k}{a^{2}}  \right)
\right].
\end{equation}
and
\begin{eqnarray*}
V=\left[\left(T_{0}-\sqrt{T_{0}(T_{0}-1)}\right)(1+w_{m})-w_{m}
\right]\rho_{m}-\frac{1}{8\pi
G}\left[\left(1-T_{0}+\sqrt{T_{0}(T_{0}-1)}\right)\dot{H}+3H^{2}
\right.
\end{eqnarray*}
\begin{equation}
\left.
+2\left(T_{0}-\sqrt{T_{0}(T_{0}-1)}+2\right)\frac{\xi-1}{T_1}H+\left(2T_{0}-
2\sqrt{T_{0}(T_{0}-1)}+1\right)\frac{k}{a^{2}} \right].
\end{equation}

$\bullet{}$ For emergent scenario, we get the expressions for
$\phi$, $T$ and $V$ as

\begin{equation}
\phi=\left(\frac{T_{0}-1}{T_{0}} \right)^{\frac{1}{4}}
\int\left[-\frac{(1+w_{m})\rho_{0}a_{0}^{-3(1+w_{m}+\delta)}}{\left(\lambda+e^{\mu
T_1}\right)^{3n(1+w_{m}+\delta)}}
 +\frac{1}{4\pi G} \left\{ -\frac{n\lambda\mu^{2}
e^{\mu T_1}}{ \left(\lambda+e^{\mu
T_1}\right)^{2}}+\frac{(\xi-1)n\mu e^{\mu T_1}}{T_1(\lambda+e^{\mu
T_1})} +\frac{k~a_{0}^{-2}}{ \left(\lambda+e^{\mu T_1}\right)^{2n}}
\right\} \right]^{\frac{1}{2}} dT_1
\end{equation}

\begin{equation}
T=\sqrt{T_{0}(T_{0}-1)}\left[-\frac{(1+w_{m})\rho_{0}a_{0}^{-3(1+w_{m}+
\delta)}}{\left(\lambda+e^{\mu T_1}\right)^{3n(1+w_{m}+\delta)}}
 +\frac{1}{4\pi G} \left\{ -\frac{n\lambda\mu^{2}
e^{\mu T_1}}{ \left(\lambda+e^{\mu
T_1}\right)^{2}}+\frac{(\xi-1)n\mu e^{\mu t}}{T_1(\lambda+e^{\mu
T_1})} +\frac{k~a_{0}^{-2}}{ \left(\lambda+e^{\mu T_1}\right)^{2n}}
\right\} \right]
\end{equation}
and
\begin{eqnarray*}
V=\left[\left(T_{0}-\sqrt{T_{0}(T_{0}-1)}\right)(1+w_{m})-w_{m}
\right]
\frac{\rho_{0}a_{0}^{-3(1+w_{m}+\delta)}}{\left(\lambda+e^{\mu
T_1}\right)^{3n(1+w_{m}+\delta)}}           -\frac{1}{8\pi
G}\left[\left(1-T_{0}+\sqrt{T_{0}(T_{0}-1)}\right)\frac{n\lambda\mu^{2}
e^{\mu T_1}}{ \left(\lambda+e^{\mu T_1}\right)^{2}}
 \right.
\end{eqnarray*}
\begin{equation}
\left.  + \frac{3n^{2}\mu^{2} e^{2\mu T_1}}{ \left(\lambda+e^{\mu
T_1}\right)^{2}}
+2\left(T_{0}-\sqrt{T_{0}(T_{0}-1)}+2\right)\frac{(\xi-1)}{T_1}
\frac{n\mu e^{\mu T_1}}{ \left(\lambda+e^{\mu
T_1}\right)}+\left(2T_{0}-2\sqrt{T_{0}(T_{0}-1)}+1\right)\frac{k~a_{0}^{-2}}{
\left(\lambda+e^{\mu T_1}\right)^{2n}} \right].
\end{equation}

$\bullet{}$ For logamediate scenario, we get the expressions for
$\phi$, $T$ and $V$ as

\begin{equation}
\phi=\left(\frac{T_{0}-1}{T_{0}}
\right)^{\frac{1}{4}}\int\left[-(1+w_{m})\rho_{0}~e^{-3A(1+w_{m}+\delta)(\ln
T_1)^{\alpha}}
 +\frac{1}{4\pi G} \left\{ \frac{A\alpha}{T_1^{2}}(\ln
T_1)^{\alpha-2}(1-\alpha+\xi \ln T_1) +k~e^{-2A(\ln T_1)^{\alpha}}
\right\}\right]^{\frac{1}{2}} dT_1
\end{equation}

\begin{equation}
T=\sqrt{T_{0}(T_{0}-1)}\left[-(1+w_{m})\rho_{0}~e^{-3A(1+w_{m}+\delta)(\ln
T_1)^{\alpha}}
 +\frac{1}{4\pi G} \left\{ \frac{A\alpha}{T_1^{2}}(\ln
T_1)^{\alpha-2}(1-\alpha+\xi \ln T_1) +k~e^{-2A(\ln T_1)^{\alpha}}
\right\}\right],
\end{equation}
and
\begin{eqnarray*}
V=\left[\left(T_{0}-\sqrt{T_{0}(T_{0}-1)}\right)(1+w_{m})-w_{m}
\right]\rho_{0}~e^{-3A(1+w_{m}+\delta)(\ln
T_1)^{\alpha}}-\frac{1}{8\pi
G}\left[2\left(T_{0}-\sqrt{T_{0}(T_{0}-1)}+2\right)\frac{(\xi-1)A\alpha}{T_1^{2}}
(\ln T_1)^{\alpha-1}
 \right.
\end{eqnarray*}
\begin{equation}
\left. +\frac{3A^{2}\alpha^{2}}{T_1^{2}}(\ln T_1)^{2\alpha-2} +
\left(1-T_{0}+\sqrt{T_{0}(T_{0}-1)}\right)
\frac{A\alpha}{T_1^{2}}(\ln T_1)^{\alpha-2}(\alpha-1-\ln T_1)
+\left(2T_{0}-2\sqrt{T_{0}(T_{0}-1)}+1\right)k~e^{-2A(\ln
T_1)^{\alpha}} \right].
\end{equation}

$\bullet{}$ For intermediate scenario, we get the expressions for
$\phi$, $T$ and $V$ as

\begin{equation}
\phi=\left(\frac{T_{0}-1}{T_{0}}
\right)^{\frac{1}{4}}\int\left[-(1+w_{m})\rho_{0}~e^{-3B(1+w_{m}+\delta)T_1^{\beta}}
 +\frac{1}{4\pi G} \left\{ B\beta(\xi-\beta)T_1^{\beta-2} +k~e^{-2B T_1^{\beta}}
\right\}\right]^{\frac{1}{2}} dT_1.
\end{equation}

\begin{equation}
T=\sqrt{T_{0}(T_{0}-1)}\left[-(1+w_{m})\rho_{0}~e^{-3B(1+w_{m}+\delta)T_1^{\beta}}
 +\frac{1}{4\pi G} \left\{ B\beta(\xi-\beta)T_1^{\beta-2} +k~e^{-2B T_1^{\beta}}
\right\}\right]
\end{equation}
and
\begin{eqnarray*}
V=\left[\left(T_{0}-\sqrt{T_{0}(T_{0}-1)}\right)(1+w_{m})-w_{m}
\right]\rho_{0}~e^{-3B(1+w_{m}+\delta)T_1^{\beta}}-\frac{1}{8\pi
G}\left[\left(1-T_{0}+\sqrt{T_{0}(T_{0}-1)}\right)B\beta(\beta-1)T_1^{\beta-2}
 \right.
\end{eqnarray*}
\begin{equation}
\left. +3B^{2}\beta^{2}T_1^{2\beta-2}
+2\left(T_{0}-\sqrt{T_{0}(T_{0}-1)}+2\right)\frac{(\xi-1)}{T_1}B\beta
T_1^{\beta-1}+\left(2T_{0}-2\sqrt{T_{0}(T_{0}-1)}+1\right)k~e^{-2B
T_1^{\beta}} \right]
\end{equation}

When we consider an interacting DBI-essence dark energy, we get
decaying pattern in the $V$-$\phi$ plot for emergent and
intermediate scenarios in the figures 10 and 12. However, from
figure 11 we see an increasing plot of $V$-$\phi$ for for
interacting DBI-essence in the logamediate scenario.

\begin{figure}
\includegraphics[height=2.7in]{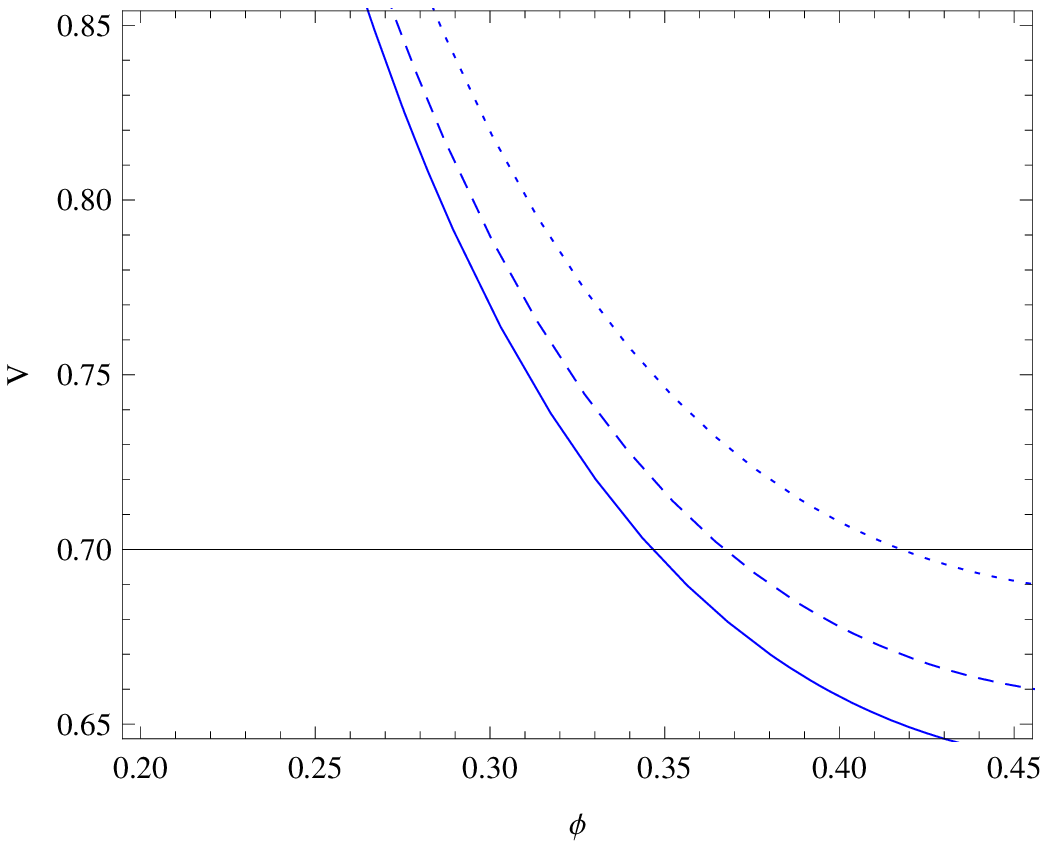}~~~~~~~~
\includegraphics[height=2.7in]{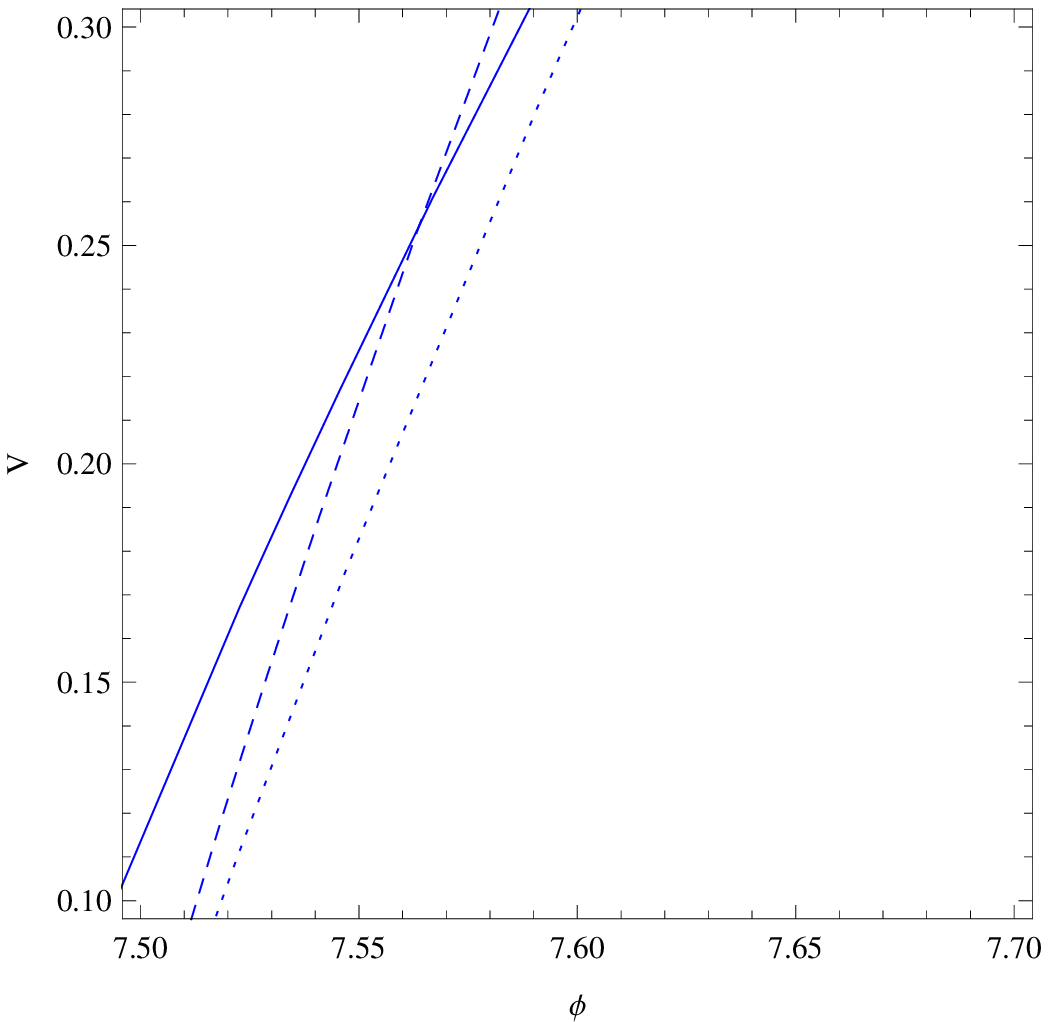}\\
\vspace{2mm}
~~~~Fig.10~~~~~~~~~~~~~~~~~~~~~~~~~~~~~~~~~~~~~~~~~~~~~~~~~~~~~~~~~~~~~~Fig.11~~\\
\vspace{6mm}
\includegraphics[height=3.0in]{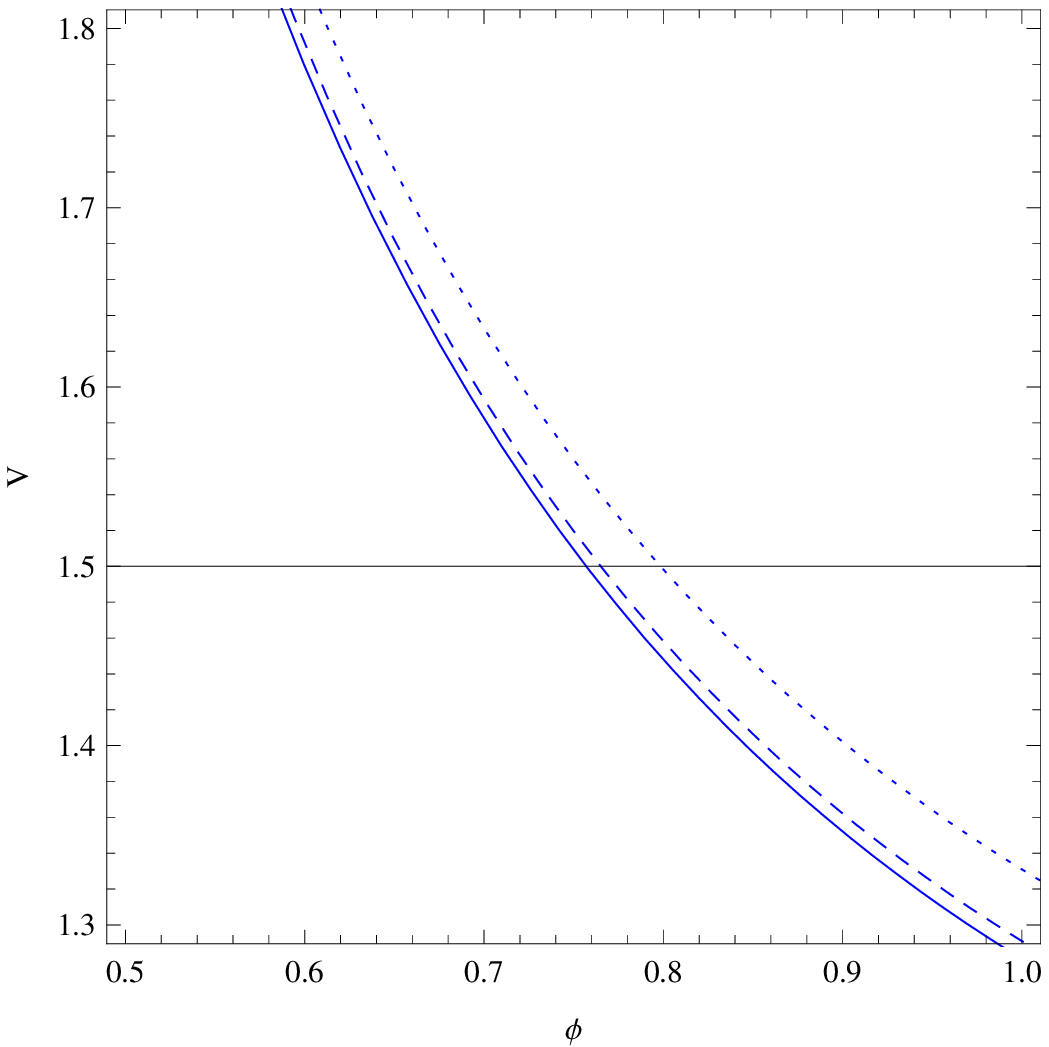}\\
\vspace{2mm} Fig.12

Figs.10-12 show the variations of $V$ against DBI field $\phi$ in
the emergent, logamediate and intermediate scenarios respectively.
Solid, dash and dotted lines represent $k=-1,+1,0$ respectively.
\end{figure}

\subsection{Hessence}

Wei et al \cite{wei} proposed a novel non-canonical complex scalar
field named ``hessence" which plays the role of quintom. In the
hessence model the so-called internal motion $\dot{\theta}$ where
$\theta$ is the internal degree of freedom of hessence plays a
phantom like role and the phantom divide transitions is also
possible. The Lagrangian density of the hessence is given by
\begin{equation}
{\cal L}_{h}=\frac{1}{2}[(\partial_{\mu}
\phi)^{2}-\phi^{2}(\partial_{\mu} \theta)^{2}]- V(\phi).
\end{equation}
The pressure and energy density for the hessence model are given by
\begin{equation}
p_{h}=\frac{1}{2}(\dot{\phi}^{2}-\phi^{2}\dot{\theta}^{2})-V(\phi),
\end{equation}
and
\begin{equation}
\rho_{h}=\frac{1}{2}(\dot{\phi}^{2}-\phi^{2}\dot{\theta}^{2})+V(\phi),
\end{equation}
with
\begin{equation}
Q=a ^{3}\phi^{2}\dot{\theta}= \text{constant,}
\end{equation}

where $Q$ is the total conserved charge, $\phi$ is the hessence
scalar field and $V$ is the corresponding potential.\\

From above we get,

\begin{equation}
\dot{\phi}^{2}-\frac{Q^{2}}{a^{6}\phi^{2}}=-(1+w_{m})\rho_{m}+\frac{1}{4\pi
G}\left(-\dot{H}+\frac{\xi-1}{T_1}H+\frac{k}{a^{2}}  \right),
\end{equation}
and
\begin{equation}
V=\frac{1}{2}(w_{m}-1)\rho_{m} +\frac{1}{8\pi
G}\left(\dot{H}+3H^{2}+\frac{5(\xi-1)}{T_1}H+\frac{2k}{a^{2}}
\right).
\end{equation}

$\bullet{}$ For emergent scenario, we get the expressions for
$\phi$ and $V$ as
\begin{equation}
\dot{\phi}^{2}-\frac{Q^{2}}{a_{0}^{6}\left(\lambda+e^{\mu
T_1}\right)^{6n}\phi^{2}}=-\frac{(1+w_{m})\rho_{0}a_{0}^{-3(1+w_{m}+\delta)}}{\left(\lambda+e^{\mu
T_1}\right)^{3n(1+w_{m}+\delta)}}
 +\frac{1}{4\pi G} \left\{ -\frac{n\lambda\mu^{2}
e^{\mu T_1}}{ \left(\lambda+e^{\mu
T_1}\right)^{2}}+\frac{(\xi-1)n\mu e^{\mu t}}{T_1(\lambda+e^{\mu
T_1})} +\frac{k~a_{0}^{-2}}{ \left(\lambda+e^{\mu T_1}\right)^{2n}}
\right\},
\end{equation}
and
\begin{equation}
V=\frac{(w_{m}-1)\rho_{0}a_{0}^{-3(1+w_{m}+\delta)}}{2\left(\lambda+e^{\mu
T_1}\right)^{3n(1+w_{m}+\delta)}}
 +\frac{1}{8\pi G} \left\{ \frac{n\mu^{2}e^{\mu
T_1}(\lambda+3ne^{\mu T_1})}{ \left(\lambda+e^{\mu
T_1}\right)^{2}}+\frac{5(\xi-1)n\mu e^{\mu T_1}}{T_1(\lambda+e^{\mu
T_1})} +\frac{2k~a_{0}^{-2}}{ \left(\lambda+e^{\mu T_1}\right)^{2n}}
\right\}.
\end{equation}

$\bullet{}$ For logamediate scenario, we get the expressions for
$\phi$ and $V$ as
\begin{equation}
\dot{\phi}^{2}-\frac{Q^{2}e^{-6A(\ln
T_1)^{\alpha}}}{\phi^{2}}=-(1+w_{m})\rho_{0}~e^{-3A(1+w_{m}+\delta)(\ln
T_1)^{\alpha}}
 +\frac{1}{4\pi G} \left\{ \frac{A\alpha}{T_1^{2}}(\ln
T_1)^{\alpha-2}(1-\alpha+\xi \ln T_1) +k~e^{-2A(\ln T_1)^{\alpha}}
\right\}
\end{equation}
and
\begin{equation}
V=\frac{(w_{m}-1)\rho_{0}}{2}~e^{-3A(1+w_{m}+\delta)(\ln
T_1)^{\alpha}}
 +\frac{1}{8\pi G} \left[ \frac{A\alpha}{T_1^{2}}(\ln
T_1)^{\alpha-2}\{\alpha-1+(5\xi-6)\ln T_1 +3A\alpha(\ln
T_1)^{\alpha}\} +2k~e^{-2A(\ln T_1)^{\alpha}} \right].
\end{equation}

$\bullet{}$ For intermediate scenario, we get the expressions for
$\phi$ and $V$ as
\begin{equation}
\dot{\phi}^{2}-\frac{Q^{2}e^{-6B T_1^{\beta}}
}{\phi^{2}}=-(1+w_{m})\rho_{0}~e^{-3B(1+w_{m}+\delta)T_1^{\beta}}
 +\frac{1}{4\pi G} \left\{ B\beta(\xi-\beta)T_1^{\beta-2} +k~e^{-2B T_1^{\beta}}
\right\},
\end{equation}
and
\begin{equation}
V=\frac{(w_{m}-1)\rho_{0}}{2}~e^{-3B(1+w_{m}+\delta) T_1^{\beta}}
 +\frac{1}{8\pi G} \left[B\beta T_1^{\beta-2}(5\xi+\beta+3B\beta T_1^{\beta})
+2k~e^{-2B T_1^{\beta}} \right].
\end{equation}

For interacting hessence dark energy, figure 13 shows increase in
the potential with scalar field and figures 14 and 15 show decay
in the potential with scalar field. This means the potential for
interacting hessence increases in the emergent universe and decays
in logamediate and intermediate scenarios.

\begin{figure}
\includegraphics[height=2.0in]{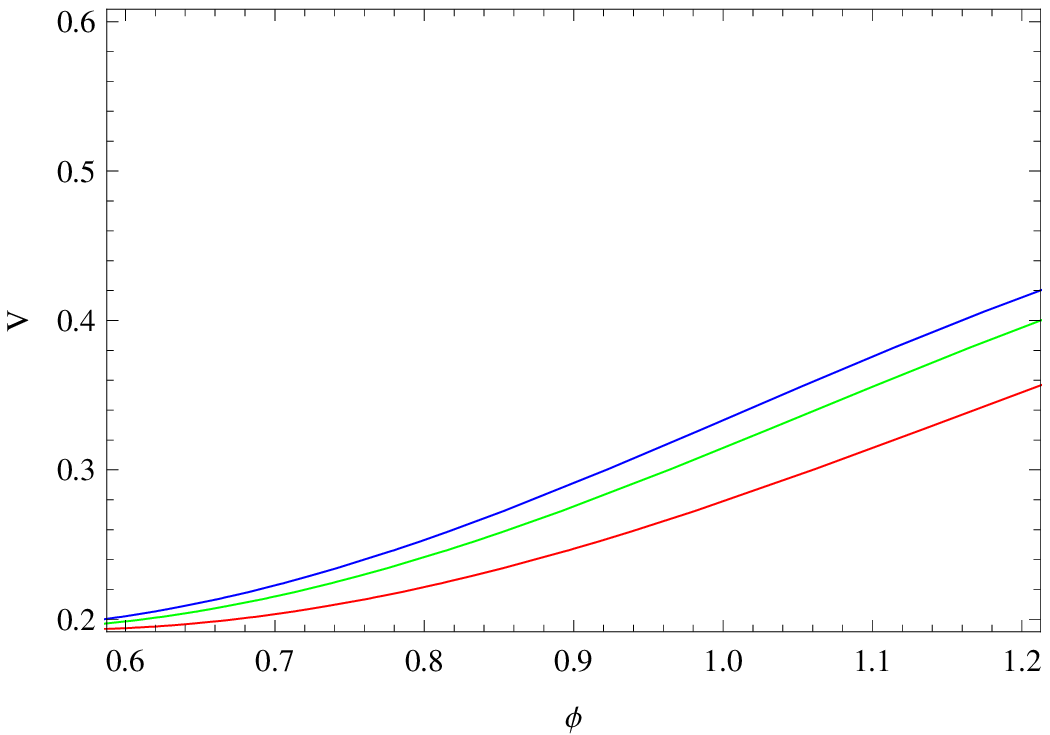}~~~~~~~~
\includegraphics[height=2.0in]{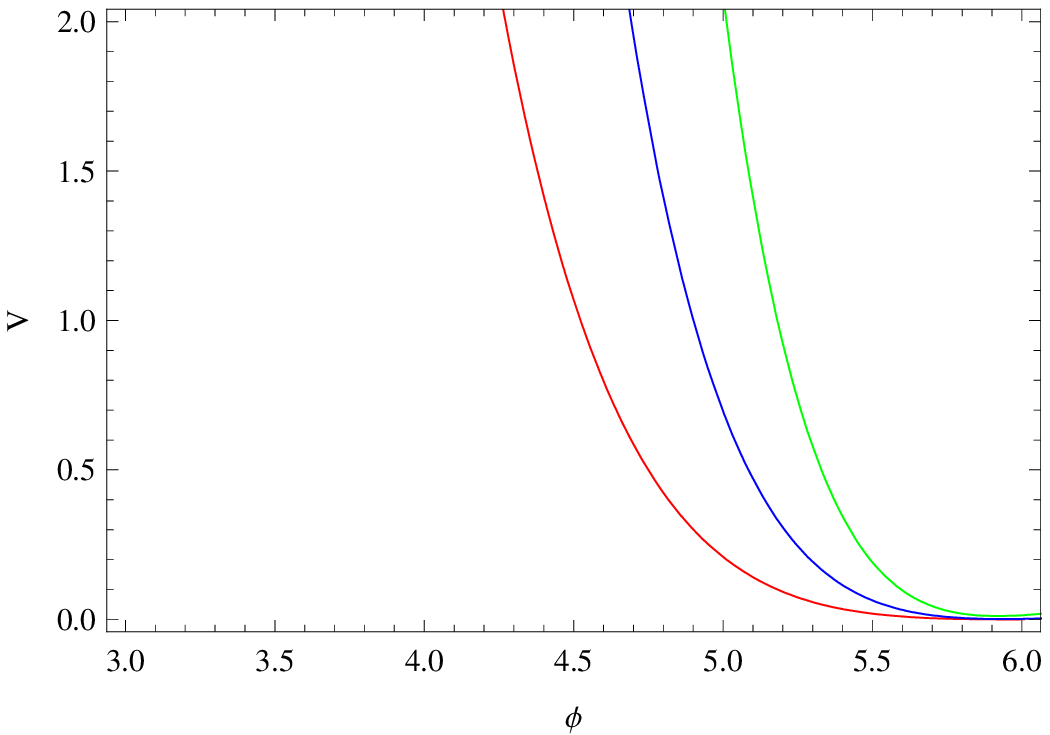}\\
\vspace{2mm}
~~~~Fig.13~~~~~~~~~~~~~~~~~~~~~~~~~~~~~~~~~~~~~~~~~~~~~~~~~~~~~~~~~~~~~~Fig.14~~\\
\vspace{6mm}
\includegraphics[height=2.5in]{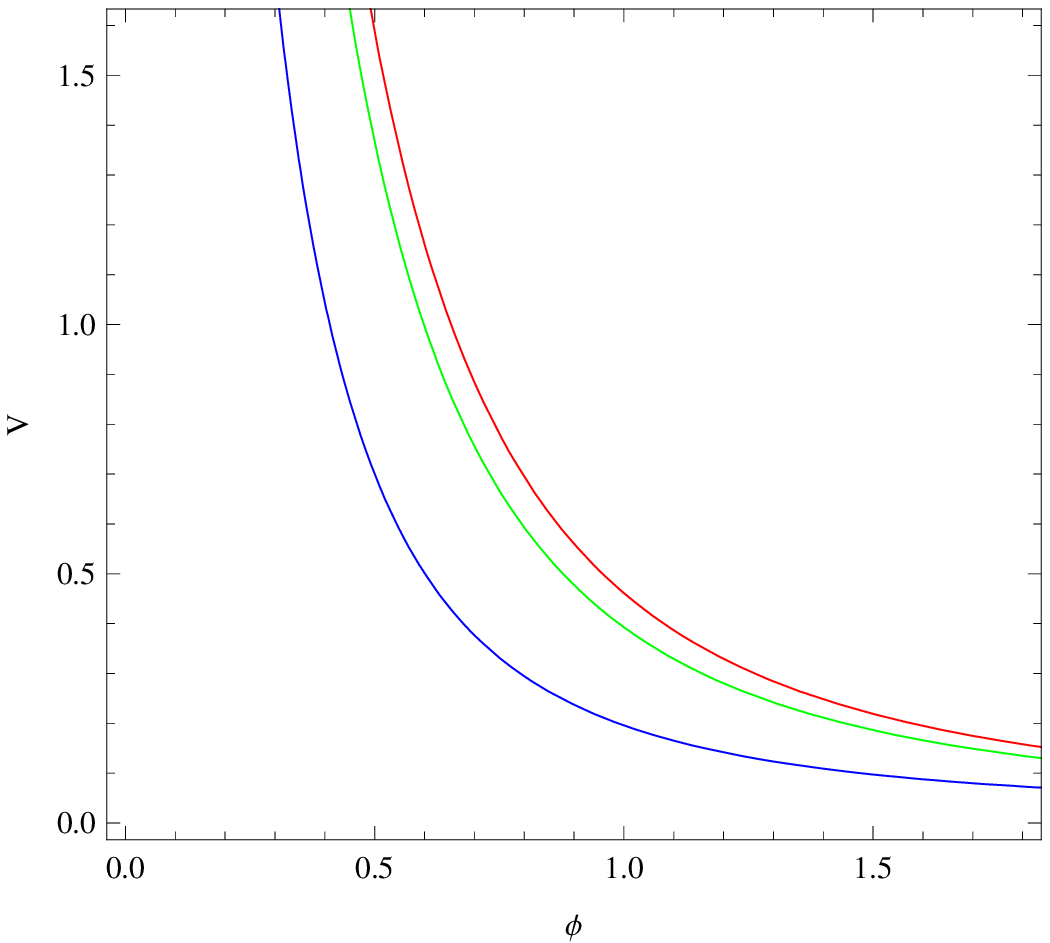}\\
\vspace{2mm} Fig.15

Figs.13-15 show the variations of $V$ against hessence field
$\phi$ in the emergent, logamediate and intermediate scenarios
respectively. Red, green and blue lines represent $k=-1,+1,0$
respectively.
\end{figure}

\subsection{Dilaton Field}

The energy density and pressure of the dilaton dark energy model are
given by \cite{Copeland}
\begin{equation}
\rho_{d}=-X+3Ce^{\lambda \phi}X^{2},
\end{equation}
and
\begin{equation}
p_{d}=-X+Ce^{\lambda \phi}X^{2},
\end{equation}
where $\phi$ is the dilaton scalar field having kinetic energy
$X=\frac{1}{2}\dot\phi^{2}$, $\lambda$ is the characteristic
length which governs all non-gravitational interactions of the
dilaton and $C$ is a positive constant.

We get,

\begin{equation}
\phi=\int \left[\frac{1}{2}(3w_{m}-1)\rho_{m} +\frac{3}{8\pi
G}\left(\dot{H}+2H^{2}+\frac{3(\xi-1)}{T_1}H+\frac{k}{a^{2}} \right)
\right]^{\frac{1}{2}}dT_1.
\end{equation}

$\bullet{}$ For emergent scenario, we have

\begin{equation}
\phi=\int
\left[\frac{(3w_{m}-1)\rho_{0}a_{0}^{-3(1+w_{m}+\delta)}}{2\left(\lambda+e^{\mu
T_1}\right)^{3n(1+w_{m}+\delta)}}
 +\frac{3}{8\pi G} \left\{ \frac{n\mu^{2}e^{\mu
T_1}(\lambda+2ne^{\mu T_1})}{ \left(\lambda+e^{\mu
T_1}\right)^{2}}+\frac{3(\xi-1)n\mu e^{\mu T_1}}{T_1(\lambda+e^{\mu
T_1})} +\frac{k~a_{0}^{-2}}{ \left(\lambda+e^{\mu T_1}\right)^{2n}}
\right\} \right]^{\frac{1}{2}} dT_1.
\end{equation}

$\bullet{}$ For logamediate scenario, we get
\begin{eqnarray*}
\phi=\int \left[\frac{3}{8\pi G} \left\{ \frac{A\alpha}{T_1^{2}}(\ln
T_1)^{\alpha-2}(\alpha-1+(3\xi-4)\ln T_1+2A\alpha(\ln T_1)^{\alpha})
+k~e^{-2A(\ln T_1)^{\alpha}} \right\} \right.
\end{eqnarray*}
\begin{equation}
\left. ~~~~~~~~~~~~~~~~~~~~~~~~~~~~~~~~~~~~~~~ +
\frac{1}{2}(3w_{m}-1)\rho_{0}~e^{-3A(1+w_{m}+\delta)(\ln
T_1)^{\alpha}} \right]^{\frac{1}{2}} dT_1
\end{equation}

$\bullet{}$ For intermediate scenario, we get

\begin{equation}
\phi=\int
\left[\frac{1}{2}(3w_{m}-1)\rho_{0}~e^{-3B(1+w_{m}+\delta)T_1^{\beta}}
 +\frac{3}{8\pi G} \left\{ B\beta(3\xi+\beta-4+2B\beta T_1^{\beta})T_1^{\beta-2}
 +k~e^{-2B T_1^{\beta}}
\right\} \right]^{\frac{1}{2}} dT_1.
\end{equation}

For interacting dilaton field, the scalar field $\phi$ always
increases with cosmic time $T_{1}$ irrespective of the scenario of
the universe we consider. This is displayed in figures 16, 17 and
18 for emergent, logamediate and intermediate scenarios
respectively.

\begin{figure}
\includegraphics[height=2.0in]{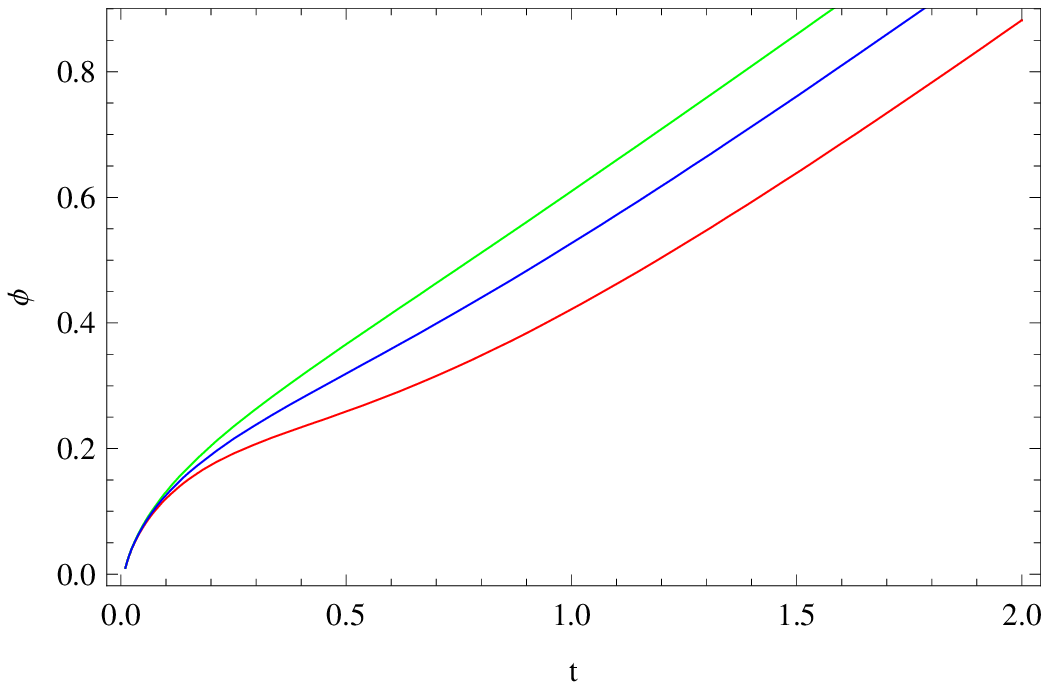}~~~~~~~~
\includegraphics[height=2.0in]{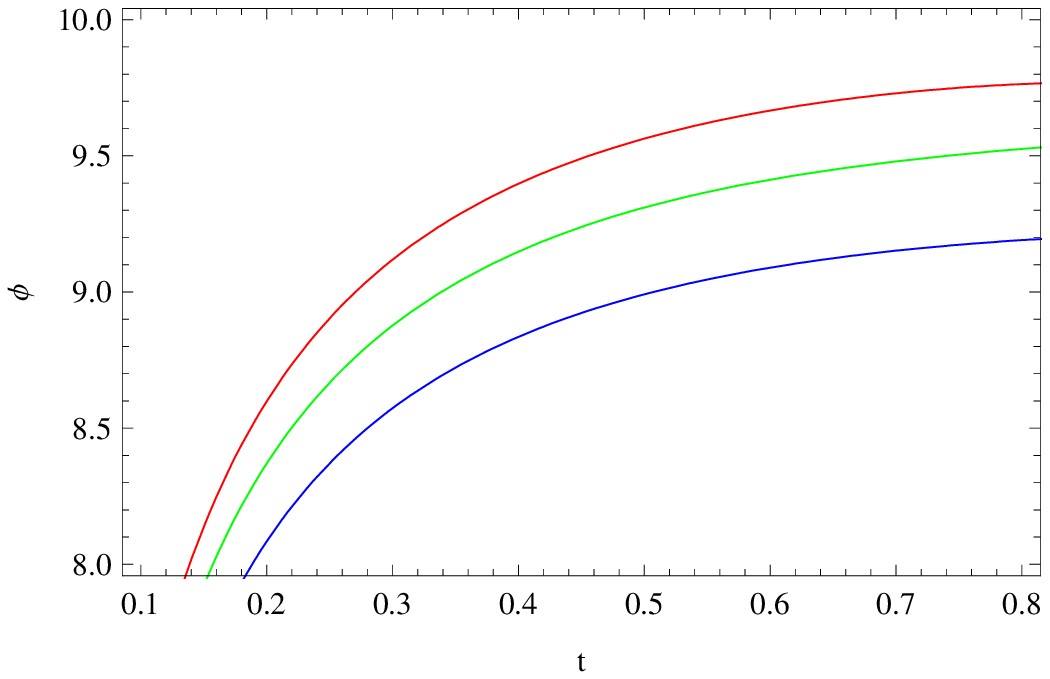}\\
\vspace{2mm}
~~~~Fig.16~~~~~~~~~~~~~~~~~~~~~~~~~~~~~~~~~~~~~~~~~~~~~~~~~~~~~~~~~~~~~~Fig.17~~\\
\vspace{6mm}
\includegraphics[height=2.0in]{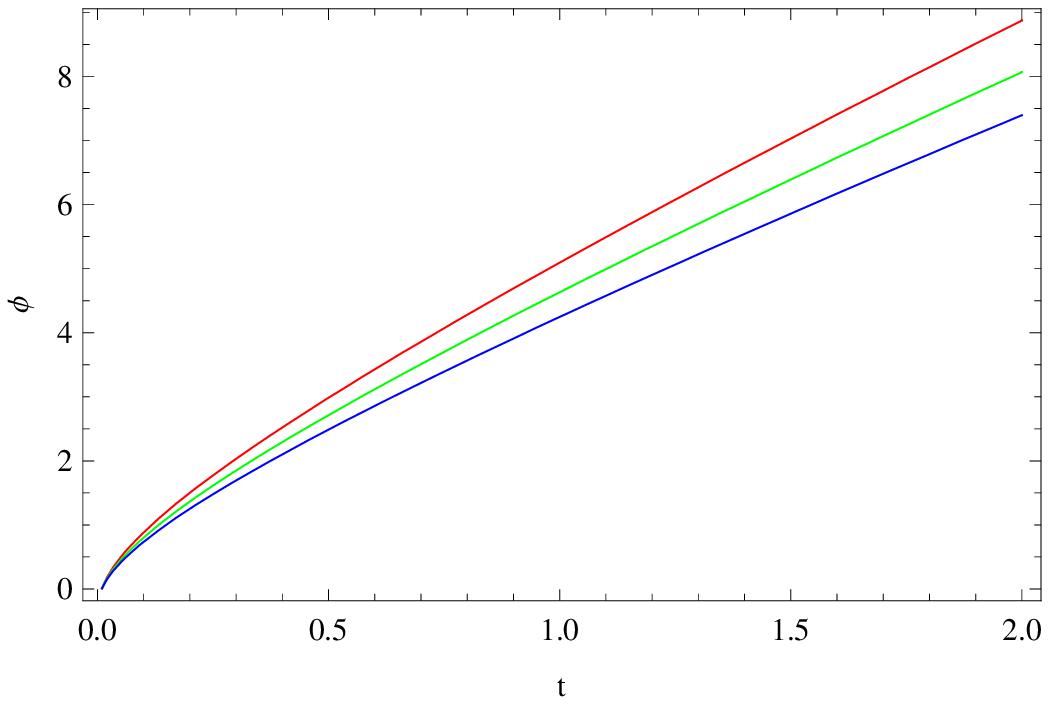}\\
\vspace{2mm} Fig.18

Figs.16-18 show the variations of dilaton field $\phi$ against time
$T_1$ in the emergent, logamediate and intermediate scenarios
respectively. Red, green and blue lines represent $k=-1,+1,0$
respectively.
\end{figure}

\subsection{Yangs-Mills Dark Energy}

Recent studies suggest that Yang-Mills field can be considered as a
useful candidate to describe the dark energy as in the normal scalar
models the connection of field to particle physics models has not
been clear so far and the weak energy condition cannot be violated
by the field. In the effective Yang Mills Condensate (YMC) dark
energy model, the effective Yang-Mills field Lagrangian is given by
\cite{Yang},
\begin{equation}
{\cal{L}}_{_{YMC}}=\frac{1}{2}bF(\ln\left|\frac{F}{K^2}\right|-1),
\end{equation}
where $K$ is the re-normalization scale of dimension of squared
mass, $F$ plays the role of the order parameter of the YMC where
$F$ is given by,
$F=-\frac{1}{2}F^{a}_{\mu\nu}F^{a\mu\nu}=E^2-B^2$. The pure
electric case we have, $B=0 ~~i.e. F = E^2$.\\

From the above Lagrangian we can derive the energy density and the
pressure of the YMC in the flat FRW spacetime as
\begin{equation}
\rho_{y}=\frac{1}{2}(y+1)b E^2,
\end{equation}
and
\begin{equation}
p_{y}=\frac{1}{6}(y-3)b E^2,
\end{equation}
where $y$ is defined as,
\begin{equation}
y=\ln\left|\frac{E^2}{K^2}\right|.
\end{equation}

We get,

\begin{equation}
E^{2}=\left[\frac{1}{2b}(3w_{m}-1)\rho_{m} +\frac{3}{8\pi
Gb}\left(\dot{H}+2H^{2}+\frac{3(\xi-1)}{T_1}H+\frac{k}{a^{2}}
\right) \right].
\end{equation}

$\bullet{}$ For emergent scenario, we have

\begin{equation}
E^{2}=
\left[\frac{(3w_{m}-1)\rho_{0}a_{0}^{-3(1+w_{m}+\delta)}}{2b\left(\lambda+e^{\mu
T_1}\right)^{3n(1+w_{m}+\delta)}}
 +\frac{3}{8\pi bG} \left\{ \frac{n\mu^{2}e^{\mu
T_1}(\lambda+2ne^{\mu T_1})}{ \left(\lambda+e^{\mu
T_1}\right)^{2}}+\frac{3(\xi-1)n\mu e^{\mu T_1}}{T_1(\lambda+e^{\mu
T_1})} +\frac{k~a_{0}^{-2}}{ \left(\lambda+e^{\mu T_1}\right)^{2n}}
\right\} \right].
\end{equation}

$\bullet{}$ For logamediate scenario, we get
\begin{eqnarray*}
E^{2}= \left[\frac{3}{8\pi bG} \left\{ \frac{A\alpha}{T_1^{2}}(\ln
T_1)^{\alpha-2}(\alpha-1+(3\xi-4)\ln T_1+2A\alpha(\ln T_1)^{\alpha})
+k~e^{-2A(\ln T_1)^{\alpha}} \right\} \right.
\end{eqnarray*}
\begin{equation}
\left. ~~~~~~~~~~~~~~~~~~~~~~~~~~~~~~~~~~~~~~~ +
\frac{1}{2b}(3w_{m}-1)\rho_{0}~e^{-3A(1+w_{m}+\delta)(\ln
T_1)^{\alpha}} \right].
\end{equation}

$\bullet{}$ For intermediate scenario, we get

\begin{equation}
E^{2}=
\left[\frac{1}{2b}(3w_{m}-1)\rho_{0}~e^{-3B(1+w_{m}+\delta)T_1^{\beta}}
 +\frac{3}{8\pi bG} \left\{ B\beta(3\xi+\beta-4+2B\beta T_1^{\beta})T_1^{\beta-2}
  +k~e^{-2B T_1^{\beta}}
\right\} \right].
\end{equation}

When we consider Yang-Mills dark energy, we find that $E^{2}$ is
always increasing with cosmic time $T_{1}$. This is displayed in
figures 19, 20 and 21 for emergent, logamediate and intermediate
scenarios respectively.

\begin{figure}
\includegraphics[height=2.0in]{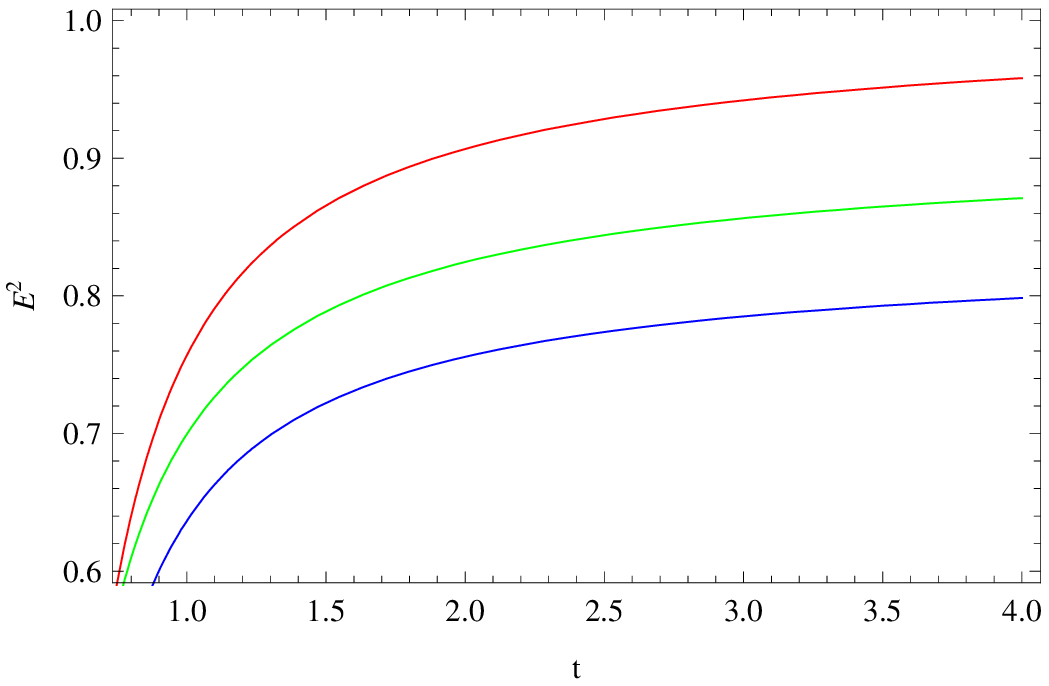}~~~~~~~~
\includegraphics[height=2.0in]{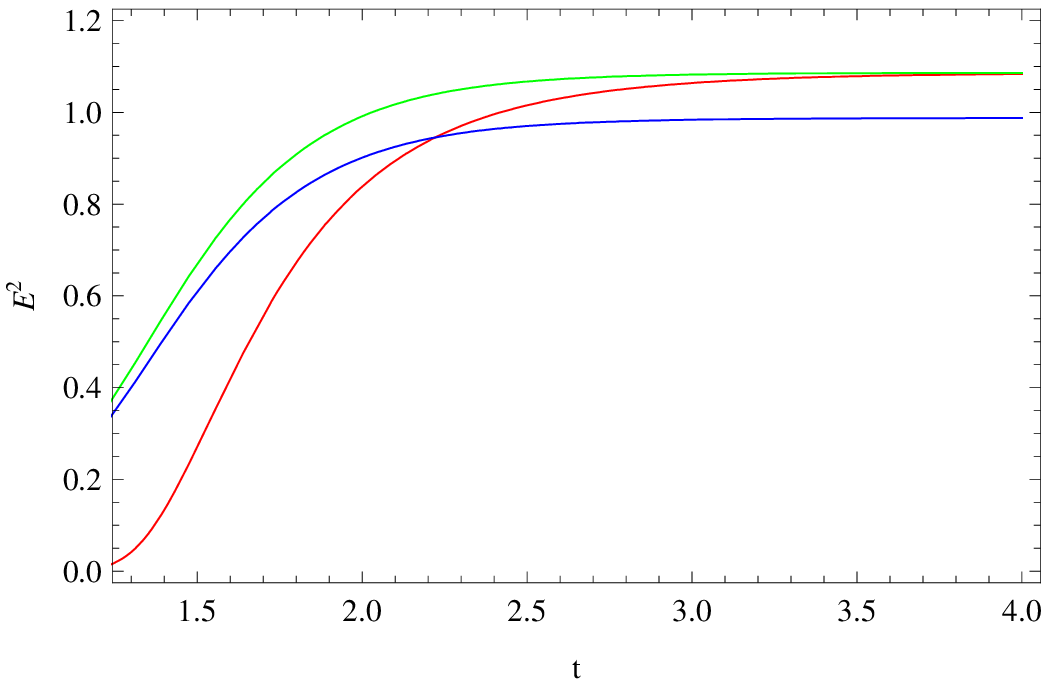}\\
\vspace{2mm}
~~~~Fig.19~~~~~~~~~~~~~~~~~~~~~~~~~~~~~~~~~~~~~~~~~~~~~~~~~~~~~~~~~~~~~~Fig.20~~\\
\vspace{6mm}
\includegraphics[height=2.0in]{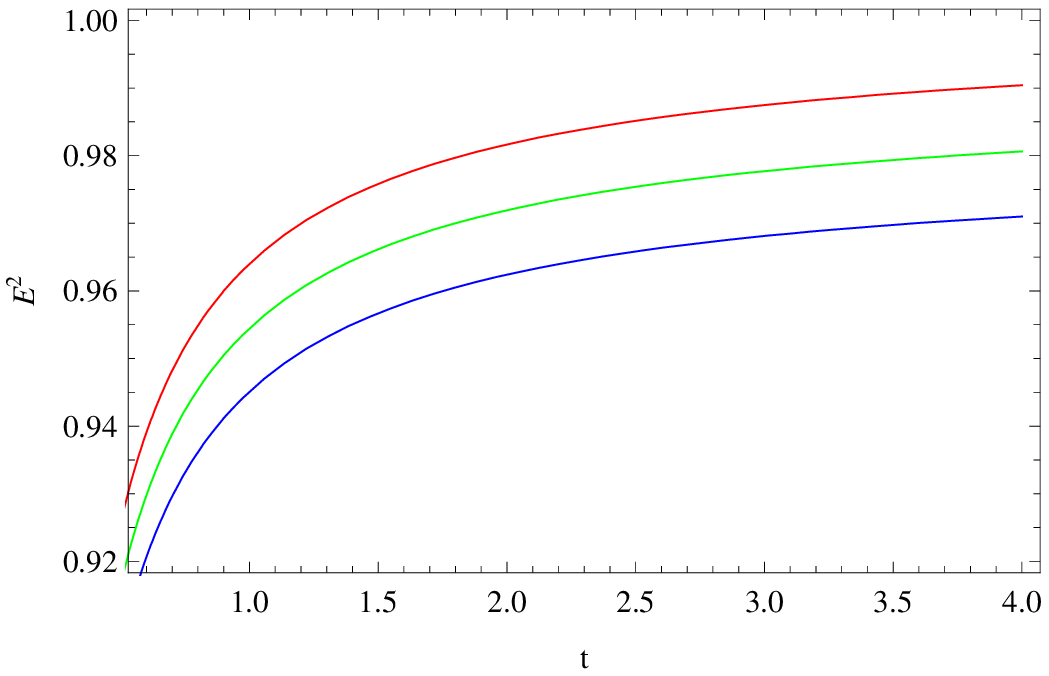}\\
\vspace{2mm} Fig.21

Figs.19-21 show the variations of $E^{2}$ against time $T_1$ in the
emergent, logamediate and intermediate scenarios respectively. Red,
green and blue lines represent $k=-1,+1,0$ respectively.
\end{figure}

\section{Conclusion}

This paper is dedicated to the study of reconstruction of scalar
fields and their potentials in a newly developed model of
Fractional Action Cosmology by Rami \cite{Rami}. The fields that
we used are quintessence, phantom, tachyonic, k-essence,
DBI-essence, Hessence, dilaton field and Yang-Mills field. We
assumed that these fields interact with the matter. These fields
are various options to model dark energy which is varying in
density and pressure, so called variable dark energy. Different
field models possess various advantages and disadvantages. The
reconstruction of the field potential involves solving the
Friedmann equations in the FAC model with the standard energy
densities and pressures of the fields, thereby solving for the
field and the potential. For simplicity, we expressed these
complicated expressions explicitly in time dependent form. We
plotted these expressions in various figures throughout the paper.

In plotting the figures for various scenarios, we choose the
following values: Emergent scenario: $\xi=.6$, $n=4$,
$\lambda=8$,$\mu=.4$, $a_0=.7$, $G=1$ (all DE models);
Logamediate: $\xi=.6$, $\alpha=3$, $A=5$, $G=1$ (all DE models);
Intermediate: $\xi=.6$, $\beta=.4$, $B=2$, $G=1$ (all DE models).
Moreover in all cases $\delta=.05$, $w_m=.01$. In figures 1 to 3,
we show the variations of $V$ against $\phi$ in the emergent,
logamediate and intermediate scenarios respectively for phantom
and quintessence field. In the first two cases, the potential
function is a decreasing function of the field. For the
quintessence field, the potential is almost constant while for the
phantom field, the potential increases for different field values.
Figures (4-6) show the variations of $V$ against $\phi$ in the
emergent, logamediate and intermediate scenarios respectively for
the tachyonic field. In figure 4, the $V$-$\phi$ plot for normal
tachyon and phantom tachyon models of dark energy is presented for
emergent scenario of the universe. Potential of normal tachyon
exhibits decaying pattern.  However, it shows increasing pattern
for phantom tachyonic field $\phi$. It happens irrespective of the
curvature of the universe. In the logamediate scenario (figure 5)
the potentials for normal tachyon and phantom tachyon exhibit
increasing and decreasing behavior respectively with increase in
the scalar field $\phi$. From figure 6 we see a continuous decay
in the potential for normal tachyonic field in the intermediate
scenario. However, in this scenario, the behavior of the potential
varies with the curvature of the universe characterized by
interacting phantom tachyonic field. For $k=-1,~1$, the potential
increases with phantom tachyonic field and for $k=0$, it decays
after increasing initially.

Similarly figures (7-9) show the reconstructed potentials for the
k-essence field. We have seen that for interacting k-essence the
potential $V$ always decreases with increase in the scalar field
$\phi$ in all of the three scenarios and it happens for open,
closed and flat universes. When we consider an interacting
DBI-essence dark energy, we get decaying pattern in the $V$-$\phi$
plot for emergent and intermediate scenarios in the figures 10 and
12. However, from figure 11 we see an increasing plot of
$V$-$\phi$ for for interacting DBI-essence in the logamediate
scenario. For interacting hessence dark energy, figures 13 shows
increase in the potential with scalar field and figures 14 and 15
show decay in the potential with scalar field. This means the
potential for interacting hessence increases in the emergent
universe and decays in logamediate and intermediate scenarios.
Figures (16-18) discuss the dilaton field while figures (19-21)
show the behavior of the Yang-Mills field in the FAC. For
interacting dilaton field, the scalar field $\phi$ always
increases with cosmic time $T_{1}$ irrespective of the scenario of
the universe and when we consider Yang-Mills dark energy, we find
that $E^{2}$ in always increasing with cosmic time $T_{1}$.

\end{document}